\documentclass{emulateapj}

\newcommand{\B}{$m_{\rm F435W}$}
\newcommand{\V}{$m_{\rm F555W}$}
\newcommand{\I}{$m_{\rm F814W}$}
\newcommand{\BI}{$m_{\rm F435W} - m_{\rm F814W}$}

\newcommand{\VI}{$m_{\rm F555W} - m_{\rm F814W}$}

\newcommand{\FeH}{[Fe/H]}

\newcommand{\afe}{[$\alpha$/Fe]}

\newcommand{\picplace}[1]{\vbox{\hrule\@height 0.4pt\@width\hsize
\hbox to\hsize{\vrule\@width 0.4pt\@height#1\hfil
\vrule\@width 0.4pt\@height#1}\hrule\@height 0.4pt\@width\hsize}}

\slugcomment{The Astronomical Journal, in press}

\shorttitle{The Wide MSTO of NGC~1846 in the LMC}
\shortauthors{Goudfrooij et al.}

\begin{document}


\title{Population Parameters of Intermediate-Age Star Clusters in 
  the \\ Large Magellanic Cloud. I. NGC 1846 and its Wide Main Sequence
  Turnoff\altaffilmark{1}}  


\author{Paul Goudfrooij\altaffilmark{2}, Thomas H. Puzia\altaffilmark{3},
  Vera Kozhurina-Platais\altaffilmark{2}, and Rupali Chandar\altaffilmark{4}}

\altaffiltext{1}{Based on observations with the NASA/ESA {\it Hubble
    Space Telescope}, obtained at the Space Telescope Science
  Institute, which is operated by the Association of Universities for
  Research in Astronomy, Inc., under NASA contract NAS5-26555} 
\altaffiltext{2}{Space Telescope Science Institute, 3700 San Martin
  Drive, Baltimore, MD 21218; goudfroo@stsci.edu, verap@stsci.edu} 
\altaffiltext{3}{Herzberg Institute of Astrophysics, 5071 West Saanich Road,
  Victoria, BC V9E 2E7, Canada; puziat@nrc.ca}
\altaffiltext{4}{Department of Physics and Astronomy, The University of Toledo,
  2801 West Bancroft Street, Toledo, OH 43606; rupali.chandar@utoledo.edu} 











\begin{abstract}
The Advanced Camera for Surveys on board the {\it Hubble Space
Telescope\/} has been used to obtain deep, high-resolution images
of the intermediate-age star cluster NGC~1846 in the Large Magellanic Cloud. 
We present new color-magnitude diagrams (CMDs) based on F435W, F555W, and F814W
imaging. We test the previously observed broad main sequence turnoff region for
`contamination' by field stars and (evolved) binary star systems. We find that while
these impact the number of objects in this region, none can fully account for
the large color spread. Our results therefore solidify the recent finding 
that stars in the main sequence turnoff region of this
cluster have a large spread in color which is unrelated to measurement
errors or contamination by field stars, and likely due to a $\sim300$~Myr range
in the ages of cluster stars.
An unbiased estimate of the stellar density distribution across the main
sequence turnoff region shows that the spread is fairly continuous rather than
strongly bimodal, as suggested previously. We fit the CMDs with several
different sets of theoretical isochrones, and determine systematic
uncertainties for population parameters when derived using any one set of
isochrones. We note a degeneracy between age and [$\alpha$/Fe],
which can be lifted by matching the shape (curvature) of the full red giant
branch in the CMD.  
We find that stars in the upper part of the main sequence turnoff region are
more centrally concentrated than those in any other region, including more
massive red giant branch and asymptotic giant branch stars.
We consider several possible formation scenarios which account for the
unusual features observed in the CMD of NGC~1846.
\end{abstract}




\section{Introduction}              \label{s:intro}

An accurate knowledge of stellar populations of ``intermediate'' age 
($\approx$\,1\,--\,3 Gyr) is important within the context of several
currently hot topics in astrophysics.  
Intermediate-age stars typically dominate the emission observed from
galaxies at high redshift \citep[e.g.,][]{vdwel+06}. Furthermore, star clusters 
in this age range are critical
for testing predictions of the dynamical evolution of star clusters 
\citep[e.g.,][]{goud+07}, and for understanding the evolution
of intermediate-mass stars. The Large Magellanic Cloud (LMC) hosts a rich
system of intermediate-age star clusters. The first surveys dedicated to studying
properties of these clusters were based on integrated colors
\citep[e.g.,][]{swb80}. These studies led to an empirical and homogeneous age
scale based on the so-called ``$S$ parameter''
\citep{elsfal85,elsfal88,gira+95,pess+08}, which describes the position of the
cluster in the integrated $(U\!-\!B)$ versus $(B\!-\!V)$ color-color
diagram. While a number of more recent studies of such 		
intermediate-age clusters have determined ages more
directly by fitting the location of the main sequence turnoff (MSTO) region
with model isochrones by means of color-magnitude diagrams
\citep[CMD; e.g.,][]{migh+98,bert+03,kerb+07,mucc+07,mack+08}, such age
determinations are still rather sparse and often dependent on the stellar
model being used. 
It is important to
obtain more CMD-based ages and metallicities of intermediate-age clusters and
to study any systematic uncertainties related to the choice of any  particular
stellar model.  

\citet{mack+08} recently used deep, HST/ACS images to study the 
intermediate age LMC star cluster NGC~1846 (plus two other clusters from 
our program GO-10595; PI: Goudfrooij). They found evidence for a wide 
range in turnoff colors in a fairly bimodal distribution, and 
concluded that these are due to two bursts of star formation separated 
by $\approx 0.3\times10^9$~yr. 
Recently, a number of star clusters, primarily massive old globular clusters
in the Milky Way have also been shown to have unusual color-magnitude diagrams
(CMDs),  suggestive of multiple stellar populations with variations 		
in either age or abundance. This includes $\omega$\,Cen
\citep[e.g.,][]{norr+96,hilric00,bedi+04,vill+07}, NGC~2808 \citep{piot+07},
NGC~1851 \citep{milo+08a}, and NGC~6388 \citep{piot08}.
These globular clusters are among the most massive known in our Galaxy, and
several are believed to be the remnant nuclei of stripped dwarf galaxies. 
NGC~1846 is roughly an order of magnitude less massive than
known Galactic clusters with multiple populations, and it seems 
unlikely to be the stripped nucleus of a cannibalized dwarf galaxy since it is
located 
in the outskirts of the LMC, which is a dwarf irregular galaxy itself. 

In this paper we conduct a more detailed investigation of NGC~1846 
in the LMC than that presented in \citet{mack+08}, including: The effects of
photometric completeness and binary star evolution, 
new techniques for assessing background contamination, 
and the radial distribution of cluster stars at different
evolutionary phases. We also investigate how well different sets of isochrones
and variations in [$\alpha/$Fe] 
fare when compared with the observations. We accomplish this by using the {\it
  ePSF} photometric technique   \citep{andkin06,ande+08a,ande+08b} to
construct the cluster CMD, while Mackey et al.\ used DOLPHOT \citep{dolp00}
for their analysis. Our results support the general conclusion of
\citet{mack+08}  that there is a large spread in the main sequence turnoff
color for NGC~1846 which must be due to a range in ages, although we do find  
differences and additional relevant properties. 

The remainder of this paper is organized as follows. \S\ 
\ref{s:obs} presents the observations, \S\ \ref{s:anal} discusses
details of the stellar photometry, completeness corrections, 
the evaluation of contamination by field stars, and radial distributions 
of stars in different parts in the color-magnitude diagram of NGC~1846.
\S\ \ref{s:isofits} presents our isochrone fitting analysis which 
involves different sets of theoretical isochrones, \S\  
\ref{s:disc} discusses the results, and \S\ \ref{s:conc} presents our
main conclusions.

\section{Observations} \label{s:obs}

NGC~1846 was observed with {\it HST\/} on Jan 12th, 2006, using the
wide-field channel (WFC) of {\it ACS\/}, as part of {\it HST\/}
program 10595 (PI: Goudfrooij). We centered the cluster on one of the two
CCD chips of the ACS/WFC camera, so that the observations cover enough
radial extent to study variations with cluster radius.
Three exposures were taken in each of the
F435W, F555W, and F814W filters: Two long exposures of 340 s each and one
shorter exposure to avoid saturation of the brightest stars
(90 s, 40 s, and 8 s in F435W, F555W, and F814W respectively). The two long
exposures in each filter were spatially offset from each other by
3\farcs011 in a direction +85\fdg28 with respect to the positive X 
axis of the CCD array. This was done to move across the gap between the
two ACS/WFC CCD chips, as well as to simplify the identification and removal of
hot pixels. 

We used two types of products provided by the {\it HST\/} calibration pipeline,
namely the {\tt flt} and {\tt drz} files. The {\tt flt} files have been
bias corrected, dark subtracted, and flatfielded.
These images are
still in the WFC CCD coordinate frame, which suffers from strong geometric
distortion. The {\tt drz} files are composite images of exposures
taken using the same filter, and are useful in the context of our analysis
for two main reasons: {\it (i)\/} The {\tt drz} files were produced using
{\it Multidrizzle\/} \citep{koek+03} which corrects the images for
geometric distortion and ties the output image to the astrometric reference
frame of the {\it HST\/} guide star catalog; {\it (ii)\/} the photometric
calibration of the ACS cameras was performed on {\tt drz} images
\citep[][hereafter S05]{siri+05}. On the other hand, the resampling of the
images by the {\it drizzle\/} algorithm \citep{fruchook02} within the {\it
  Multidrizzle\/} package makes them less suited 
for accurate point-spread function (PSF)-fitting photometry. Hence, we use
the {\tt drz} files for photometric and astrometric calibration and the
{\tt flt} files for measurements.  

\section{Analysis}          \label{s:anal}

\subsection{Photometry}    \label{s:phot}

Stellar photometry was performed using PSF fitting, using the spatially
variable ``effective PSF'' {\it (ePSF)} method described in
\citet{andkin00}, and tailored for ACS/WFC data by \cite{andkin06}. A detailed
description of the application of the ePSF method to ACS/WFC data is given
in \citet{ande+08a,ande+08b}.  Small temporal variations of the PSF from one image to
another (likely due to thermal breathing and/or small changes in the focus) were dealt
with by constructing a spatially constant ``perturbation'' PSF for each
image which was added to the library {\it ePSF} for each filter. 

After fitting positions and fluxes of all stars in all individual
{\tt flt} images using the {\it ePSF}-fitting algorithm, they were
transformed into the geometrically corrected {\tt (drz)} image
reference frame using the prescriptions of \citet{andkin06}. We
selected all stars with the {\it ePSF} parameter ``PSF fit quality'' $q <
0.5$  and ``isolation index'' of 5. The latter parameter 
selects stars that have no brighter neighbors within a radius of 5
pixels. To further weed out hot pixels,
cosmic rays and spurious detections along diffraction spikes, the
geometrically corrected positions among the three images per 
filter were compared. We selected objects with coordinates matching
within a tolerance of 0.2 pixels in either axis, which eliminated the
hot pixels and cosmic ray hits effectively. The photometry, at this
stage, is given as instrumental magnitudes ($-2.5 \log_{10}
(\mbox{flux [e$^-$]})$) for each individual exposure.  

\begin{figure}
\centerline{\includegraphics[width=8.3cm]{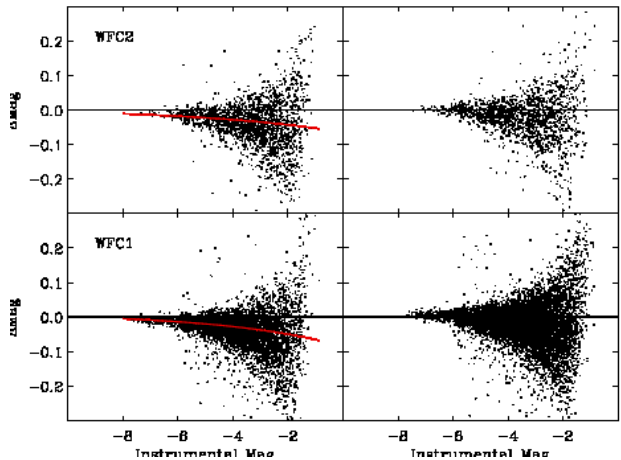}}
\caption{The difference in normalized instrumental F435W magnitude ($-2.5$ log
  (e$^-$ s$^{-1}$)) derived
  from short vs.\ long exposures as a function of normalized instrumental
  F435W magnitude of the 
  long exposures. The left panels show the observed values (affected by CTE
  loss), whereas the right panels show the result of the CTE correction. The top
  and bottom panels refer to stars on the WFC2 and WFC1 chips,
  respectively. The overplotted solid (red online) line illustrates the
  average trend in magnitude residuals. 
\label{f:ctefig1}}  
\end{figure}

\begin{figure}
\centerline{\includegraphics[width=8.3cm]{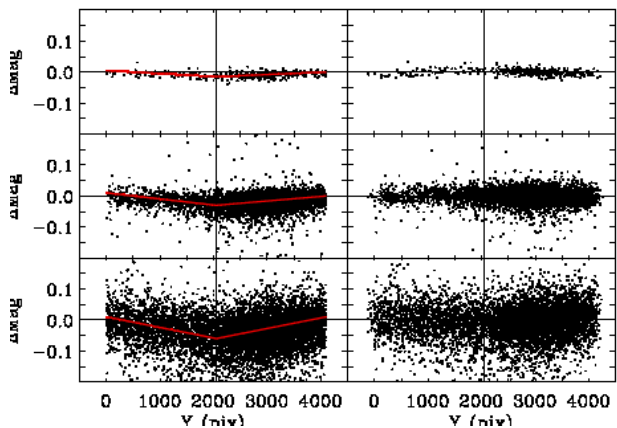}}
\caption{The difference in normalized instrumental F435W magnitude derived
  from short vs.\ long exposures as a function of $Y$ coordinate in the {\tt
    drz} reference frame. The left panels show the observed values (affected
  by CTE loss), whereas the right panels show the result of the CTE
  correction. The top row of panels refer to stars with normalized
  instrumental magnitudes $-10 \leq {\rm mag} < -6$, the middle row does so for
  $-6 \leq {\rm mag} < -4$, and the bottom row does so for $-4 \leq {\rm mag}
  < -2$. The overplotted solid (red online) line illustrates the
  average trend in magnitude residuals for each chip (WFC2 for $Y \leq 2048$
  and WFC1 for $Y > 2048$). 
\label{f:ctefig2}}  
\end{figure}

Corrections for imperfect charge transfer efficiency (CTE) of the ACS/WFC
CCDs were made following \citet{kozh+07}, using a functional form of 
CTE loss that was derived from the data itself by comparing
the photometry of stars from the short versus the long exposures. 
%
Using this technique, we find a different functional form for CTE loss
than that published by \citet{riesmack04}. The latter was found to
overcorrect the CTE loss. 
This difference is
likely due to two main reasons:\ {\it (i)\/} The \citet{riesmack04}
correction formulae were derived from sparsely populated
fields, whereas we are working with a much more crowded field. Charge traps
in crowded fields are expected to be filled by charge associated with stars
located at rows close to the read-out amplifier and hence the overall
effective CTE loss is expected to be lower than in sparse fields \citep[see,
e.g.,][]{goud+06}; {\it (ii)\/} 
\citet{riesmack04} performed aperture photometry on {\tt 
  drz} images, whereas we are using PSF-fitting photometry on {\tt flt}
files. The latter method assigns higher weights to the central pixels of stars
relative to those located in the wings of the PSF, whereas aperture
photometry assigns equal weights to every pixel within the
aperture. 
The relevance of the CTE correction is illustrated in Figs.\ \ref{f:ctefig1}
and \ref{f:ctefig2}. 
The accuracy of our CTE corrections is such that the
rms scatter of the magnitude residuals is 0.02 mag at an instrumental
magnitude of $-2$ (corresponding to \B\ = 23.8, \V\ = 24.4, and \I\ = 23.5 
in magnitude units relative to Vega). This uncertainty is smaller than the
photometric measurement errors at those magnitudes.  
After CTE correction, a final magnitude for each object was determined from
a weighted average from all three available photometric measurements in
a given filter.
A weight of $\sigma^{-2}$ was used after rejecting saturated sources in the
long exposures. 


\subsection{Photometric Calibration}

We determined final magnitudes for each star on the Vegamag photometric system
following the procedure outlined in \citet{bedin+05}. 
Briefly, instrumental magnitudes measured with the {\it ePSF} method were
compared with aperture photometry with a $0.5\arcsec$ radius of the same
stars. Our final, calibrated magnitudes were then determined using:
\begin{eqnarray}
m_{\rm filter} = & -2.5 \, \log_{10} \left( \frac{{\rm signal}\, [e^-]}{{\rm
      exptime}} \right) + {\it ZP}_{\rm filter}  \nonumber \\
  & - \Delta m_{0''\!\!\! .\,5-\infty}^{\rm filter} - \Delta m_{{\rm PSF} -
    0''\!\!\! .\,5}^{\rm filter} \nonumber
\end{eqnarray}
where the subscript ``filter'' refers to the specific filter that was used. 
The first term on the right hand side is the normalized instrumental
magnitude measured from the {\tt flt} images using {\it ePSF},
the second term is the zeropoint from S05\footnote{The S05 zeropoints 
  were recently updated for different observing dates, and are available at 
  http://www.stsci.edu/hst/acs/analysis/zeropoints.}, the 		
third term is the correction from 0\farcs5 to an infinite
aperture \cite[taken from table 1 in][]{bedin+05}, and the fourth term is the
empirically determined difference between our ePSF-fitting photometry and the
aperture photometry performed for bright isolated stars in the image.

\subsection{Color-Magnitude Diagrams}  \label{s:CMDs}

\begin{figure*}
\centerline{\includegraphics[width=0.65\textwidth]{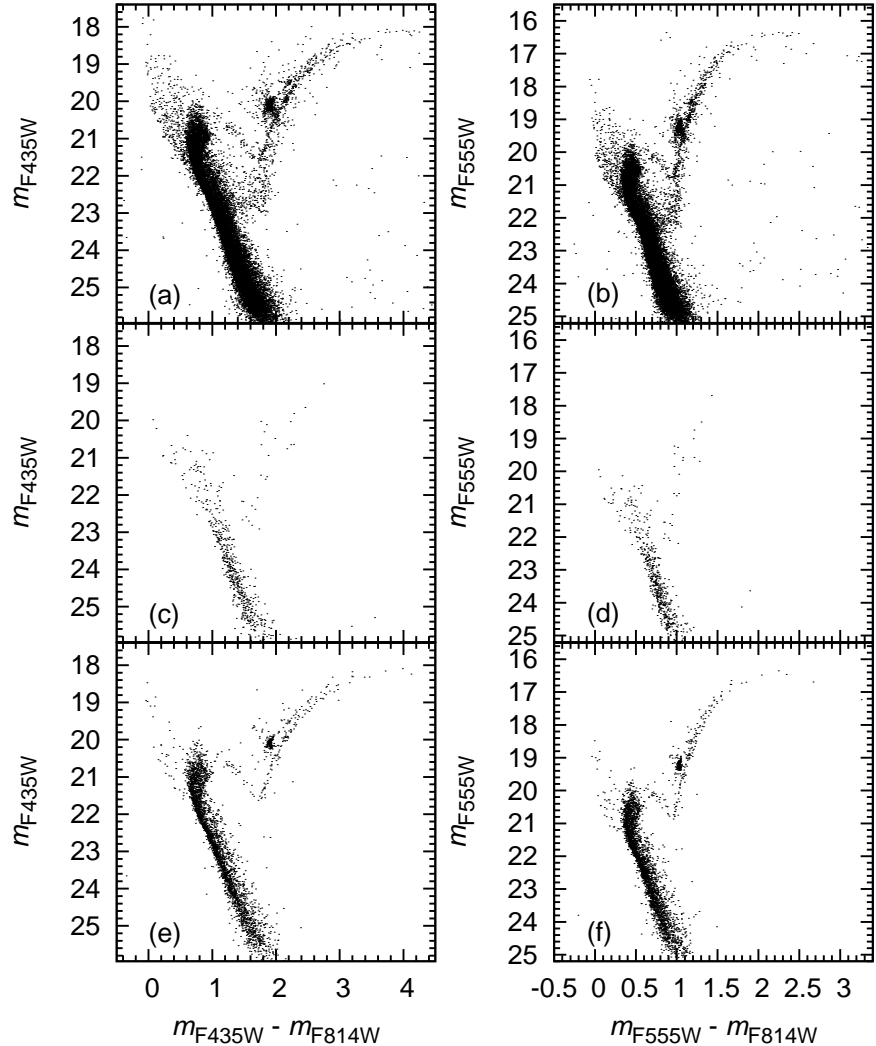}}
\caption{CMDs for the HST/ACS images of NGC~1846. The top panels (a) and (b)
  show \B\ vs. \BI\ and  \V\ vs.\ \VI\ for all 23,652 detected sources that passed our selection 
  criteria. The middle panels (c) and (d) do so for a region at radii
  125$''$\,--\,190$''$ from the center of NGC~1846 (near the corners of
  the {\it HST/ACS\/} image), and the bottom panels (e) and (f) do so for the selected
  sources within 25$''$ from the center of NGC~1846, a region with the
  same area as that shown in panels (c) and (d). 
\label{f:fullCMDs}}  
\end{figure*}

Color-magnitude diagrams (CMDs) for NGC~1846 were created for both \B\ vs.\ 
\BI\ and \V\ vs.\ \VI. All stars selected as described in \S\ \ref{s:phot}
above are plotted in panels (a) and (b) of Fig.\ \ref{f:fullCMDs}. To decrease
contamination by field stars, we also plot CMDs including only stars 
within 25 arcsec  from the center of NGC~1846 (which has reference 		
pixel coordinates $(x_c,\,y_c)  = (2225, 3096)$, see \S\ \ref{s:rad_dist} 
below) in panels (c) and (d). 
There are several features clearly visible in the CMD of NGC~1846 shown in
Fig.\ \ref{f:fullCMDs}:  
\begin{enumerate} 
\item 
The main sequence is well defined and extends over about 6 mag in \B\ and \V, and is
accompanied by an obvious sequence of unresolved binary stars
(running along the main sequence but slightly brightward and redward). The
main sequence
shows a slight change in slope near \B\ $\approx$ 22.5, flagging the transition between
radiative and convective stellar cores. Fig.\ \ref{f:fullCMDs} shows 
 that the MSTO region is fairly wide, whereas the MS of single stars 
fainter than the turnoff
is well-defined and much narrower (see especially panel
 (e)). This constitutes strong evidence for the presence of more than one
 population, which is discussed in more detail below.   
\item 
The subgiant branch (SGB) is quite well defined and seems to emerge mainly
from the upper end of the MSTO region. 
\item 
The red giant branch (RGB) is well defined and quite narrow
(see especially panels (a) and (e) of Fig.\ \ref{f:fullCMDs}). 
This strongly suggests that there is little differential reddening
across NGC~1846. Note that the significant width in color of the MSTO
region is not reflected in the RGB. 
This argues that the populations that make up the MSTO share the same 
metallicity (in terms of [Fe/H]).

The data allow one to clearly identify the so-called ``RGB bump'' at \B\ =
20.0, \BI\ = 2.1, \V\ = 19.0, and \VI\ = 1.1. This feature was first
predicted by \citet{thom67} and \citet{iben68}, and corresponds to the period
during Hydrogen shell burning when the outward-moving shell encounters
extra Hydrogen fuel left behind during the period when the
star had a convective envelope. The ``bump'' occurs as the star adjusts its
structure as it consumes this extra fuel, briefly reversing its path through
the CMD before resuming its ascent of the RGB  
\citep[see also, e.g.,][]{fusi+90,denvdb03}. 
\item 
The red clump (RC), sometimes called the Helium clump (since it represents
core Helium burning), is located at \B\ $\simeq$ 20.1, \BI\ $\simeq$
1.9, \V\ $\simeq$ 19.2, and \VI\ $\simeq$ 1.0.  
\item
The asymptotic giant branch (AGB) clump can be seen at \B\ $\simeq$ 19.5, \BI\
$\simeq$ 2.2, \V\ $\simeq$ 18.4, and \VI\ $\simeq$ 1.2. This feature
corresponds to the onset of Helium shell burning.  
\end{enumerate}

The wide MSTO region in NGC~1846 was previously reported by \citet{macbro07} and
\citet{mack+08}, who interpreted this feature as due to the presence of two 
stellar populations of
identical metallicity but different age. Our photometry shows some differences
with that of Mackey and coworkers that potentially affects the interpretation of how this
cluster formed.
These differences are discussed further in
\S\ \ref{s:disc}.  

\subsection{Completeness Corrections} \label{s:completeness}

\begin{figure*}[htb]
\centerline{\includegraphics[angle=-90,width=0.7\textwidth]{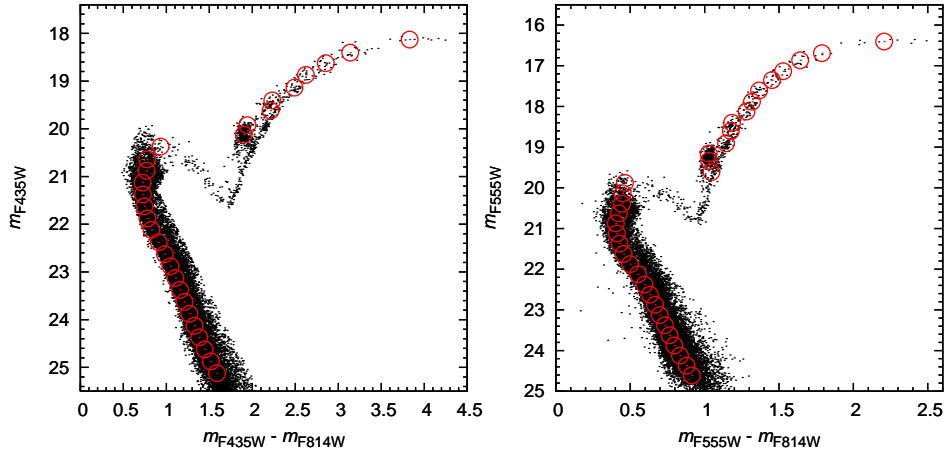}}
\caption{CMDs of stars in regions in color-magnitude space that have less than
  10\% of field star contamination (see text in \S\ \ref{s:CMDs}). Left
  panel: \B\ vs.\ \BI. Right panel: \V\ vs.\ \VI. The open red circles in the 
  CMDs depict the magnitudes and colors used for the 
  completeness fractions shown in Fig.\ \ref{f:completeness} (see \S\
  \ref{s:completeness}). 
\label{f:clusterCMDs}}  
\end{figure*}

\begin{figure*}[htb]
\epsscale{0.7}
\plotone{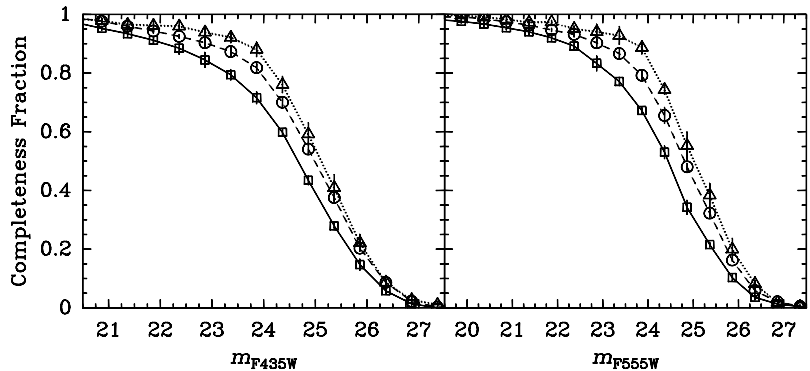}
\caption{The completeness fraction as function of magnitude and radius. Open
  squares and solid lines represent data at a radius of 5$''$, open circles
  and dashed lines represent data at a radius of 35$''$, and open triangles
  and dotted lines represent data at a radius of 125$''$. Error bars depict
  the standard deviations among 50 independent runs of the artificial star
  tests. See description in \S\ \ref{s:completeness}. 
\label{f:completeness}}  
\end{figure*}

Density distributions of stars in crowded images suffer from incompleteness.
We use the standard technique of adding artificial stars to the images
and running them through the photometric pipeline in order to quantify
incompleteness.
Specifically we use a program written by Jay
Anderson \citep[see][]{ande+08a} that uses the best-fit arrays of ePSFs for each
individual image, repeatedly adding artificial stars covering the magnitude
and color ranges found in the CMDs. 
The overall radial distribution of the inserted artificial stars followed that
of the stars in the image. 
Only 5\% of the total number of stars selected in \S\ \ref{s:phot} above
were added per run, so as not to increase the degree of crowding in the 
images significantly.
After
inserting the artificial stars, the ePSF measurement and selection procedures
were applied again to the image. An inserted star was considered recovered if
{\it (i)\/} the input and output positions agreed to within 0.3 pixel, and
{\it (ii)\/} the input and output fluxes agreed to within 0.5 mag in every
filter. Completeness fractions of stars as a function of magnitude and 
distance from the cluster center are shown in Fig.\ \ref{f:completeness} for 
stars in the cluster sequences, as depicted in Fig.\ \ref{f:clusterCMDs}. 

Completeness fractions were assigned to every individual star in a given CMD
by fitting the completeness fractions as a function $f_{\rm comp} (m,r)$ of
magnitude $m$ and radius $r$: 
\begin{equation}
f_{\rm comp} (m,r) = \frac{1}{2} 
 \left[ 1 - \frac{\alpha(r) \, (m - m_{\rm lim} (r))}{\sqrt{1 + \alpha^2 (r)\,
       (m - m_{\rm lim} (r))^2}} \right]
\end{equation}
where $\alpha (r)$ is a free scaling parameter and $m_{\rm lim} (r)$ is the magnitude
where completeness is 50\%. The functional forms of both $\alpha (r)$ and
$m_{\rm lim} (r)$ were chosen as third-order polynomials, fit to
the data using 10 bins in radius. The overall rms of the fit to the
completeness data was 0.03 for both the \B\ vs.\ \BI\ and the \V\ vs.\ \VI\
CMDs. 

\subsection{Radial Distributions} \label{s:rad_dist} 

The wide main sequence turn-off region
in  NGC~1846 suggests the existence of more than one simple population. If so,
it is important to find out whether the   
different populations have intrinsically different spatial distributions, i.e.,
differences beyond those that can be expected for simple 
(coeval) stellar populations due to dynamical evolution. A well-known example
is mass segregation due to dynamical friction which slows down stars on a time
scale that is inversely proportional to the mass of the star
\citep[e.g.,][]{sasl85,spit87,mayheg97} so that massive stars have a more
centrally concentrated distribution over time than less massive ones.  
With this in mind, we determined regions in the CMD that are strongly
dominated by stars belonging to the cluster rather than to the underlying
field in the LMC.  

\subsubsection{Surface Density Distribution}

\begin{figure}
\centerline{\includegraphics[bb=213 70 540 543,clip,angle=-90,width=8.3cm]{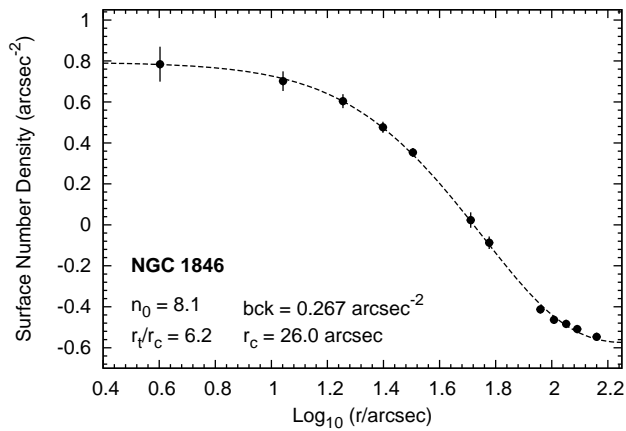}}
\caption{Radial surface density profile of NGC 1846. The points
  represent observed values. The dashed line represents the best-fit
  King model (cf.\ Eq.\ \ref{eq:King}) whose parameters are shown in 
  the legend.  
\label{f:numdensfit}}  
\end{figure}

We first determine the projected
number density of stars, taking into account the completeness corrections. The
cluster  center was determined to be at reference coordinate ($x_c$,\,$y_c$) = 
(2225,\,3096) with an uncertainty of $\pm$\,5 pixels in either axis. This was
done by creating a 2-D histogram of the pixel coordinates of stars using a bin
size of $50 \times 50$ pixels (i.e., 2\farcs5 $\times$ 2\farcs5), and then
calculating the center using a 2-D gaussian fit to an image constructed from
the surface number density values in the 2-D histogram. Note that this method
avoids biases related to the presence of bright stars near the center. The
ellipticity of NGC~1846 was found to be 0.12\,$\pm$\,0.02, which was derived
by running the task {\tt ellipse} within {\sc iraf/stsdas}\footnote{STSDAS is
  a product of the Space Telescope Science Institute, which is operated by
  AURA for NASA} on the surface number density image mentioned above.  
The area sampled by the ACS image was divided in 12 concentric elliptical 
annuli, centered on the cluster center. For annuli with radii larger than
$\sim$\,850 pixels ($\hat{=}$ 42\farcs5), care was taken to account for the
limited azimuthal coverage of the cluster in the image. Radii are
expressed in terms of the ``equivalent'' radius of the ellipse, $r = a\,
\sqrt{1-\epsilon}$ where $a$ is the semimajor axis of the ellipse and
$\epsilon$ its ellipticity.  
The radial surface number density profile was fit with a \citet{king62} model
combined with a constant background level:  
\begin{equation}
n(r) = n_0 \: \left( \frac{1}{\sqrt{1 + (r/r_c)^2}} - \frac{1}{\sqrt{1+c^2}}
 \right)^2 \; + \; {\rm bkg} 
\label{eq:King}
\end{equation}
where $n_0$ is the central surface number density, $r_c$ is the core radius,
and $c \equiv (r_t/r_c)$ is the concentration index ($r_t$ being the tidal
radius). The best-fit King model was selected using a $\chi^2$
minimization routine, and is shown in Fig.\ \ref{f:numdensfit} as well
as the individual surface number density values.  

\subsubsection{Selection of Cluster-Dominated Regions on the CMD}

\begin{figure}
\centerline{\includegraphics[bb=163 147 540 539,clip,angle=-90,width=8.3cm]{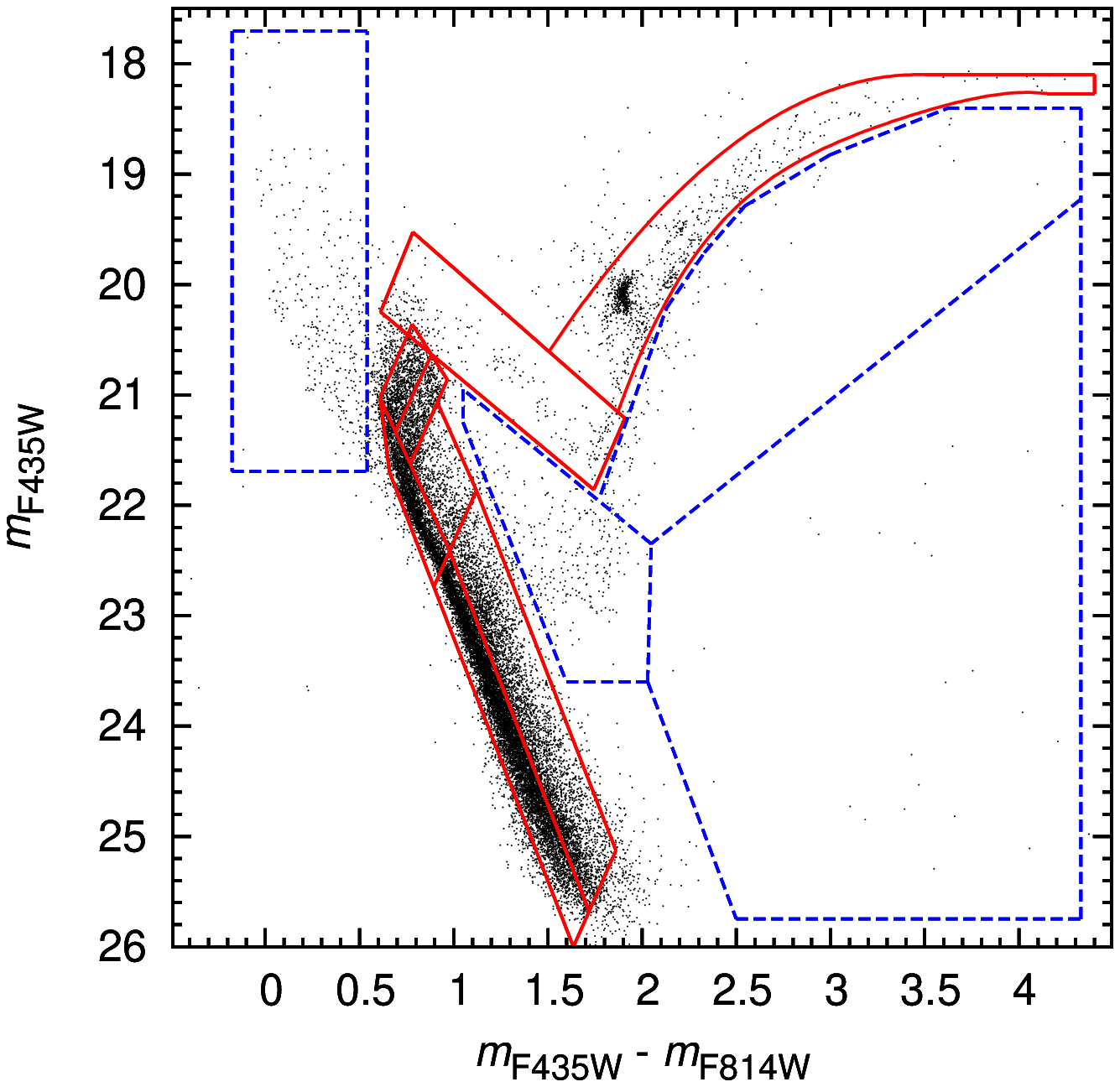}}
\caption{A copy of Fig.\ \ref{f:fullCMDs}a showing regions of
  interest. The red solid lines encompass regions in color-magnitude
  space that have less than 10\% of field star contamination and for
  which radial distributions are shown and described in Fig.\ 
  \ref{f:rad_dists}. The blue dashed lines encompass regions found to
  be heavily contaminated by field stars. See discussion in \S\ 
  \ref{s:rad_dist}.2.    		
\label{f:regions}}  
\end{figure}

The next step in establishing regions on the CMD that are strongly dominated
by cluster stars was done statistically, by comparing star surface number
densities in selection boxes on the CMD from two radial ranges: ``Inner''
stars with $\log(r) \leq 1.5$ versus ``outer'' stars with $2.0 \leq \log(r)
\leq 2.2$ (with radius $r$ in arcsec). Selection boxes for which the
``inner/outer'' surface number density ratio (after completeness correction)
exceeded the value given by the best-fit King model to {\it all} stars in the
ACS image were tagged as dominated by cluster stars. The selection boxes
themselves were created for each evolutionary sequence in the CMD (MS, RGB, He
clump, AGB), plus (bigger) boxes in the less populated areas of the CMD. For
each evolutionary sequence, a ridgeline was established by means of a spline
fit to carefully selected reference points along the sequence in
question. After subtracting the color of the ridgeline from those of the stars
in that sequence, selection boxes were divided into bins of magnitude and
color. The bin size for the MS region was chosen adaptively depending on the
local number density in the CMD, with a minimum of 0.2 mag in magnitude and
color. For the RGB and AGB, the bin size in color was chosen to properly
separate the two sequences. The selection boxes that did not correspond to the
evolutionary sequences are depicted as such on Fig.\ \ref{f:regions} with
dashed lines.  

\begin{figure*}
\centerline{\includegraphics[bb=61 308 536 565,width=0.75\textwidth]{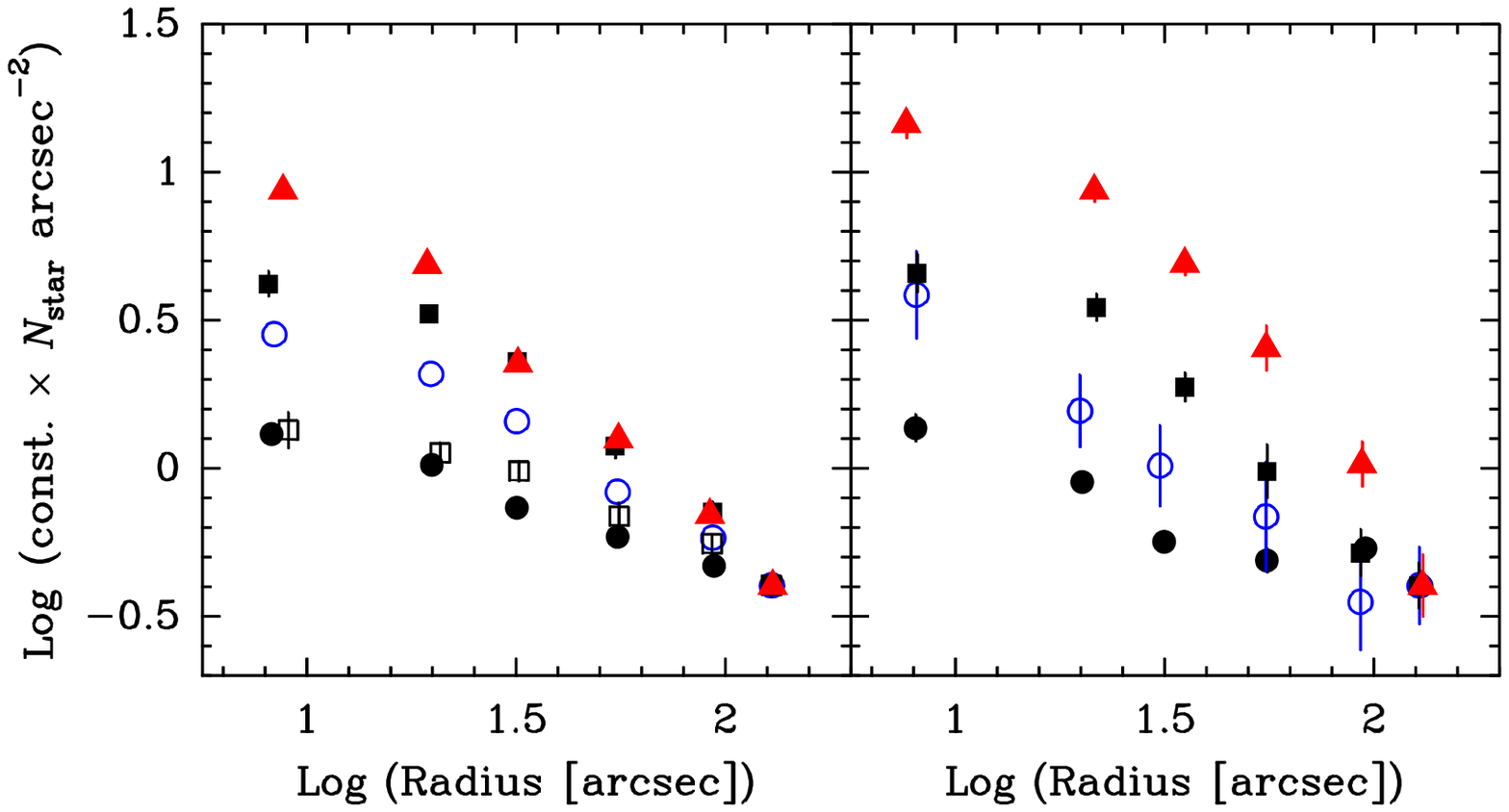}}
\caption{Radial surface number density distributions of stars in different
  regions of the CMD, as depicted in Figure \ref{f:regions} with red solid   
  lines.  {\it Left
    panel:\/} Solid circles:\ all stars in the CMD. Open circles:\ stars in
  the regions of the CMD determined to have less than 10\% contamination by
  field stars. Open squares: lower MS stars. Filled squares:\ upper MS stars. 
  Filled triangles:\ RGB and AGB stars. {\it Right panel:\/} Solid
  circles:\ stars in lower binary sequence. Open circles:\ stars in
  upper binary sequence. Filled squares:\ lower MSTO region. Filled
  triangles:\ upper MSTO region. 
  The absolute surface number density values on the Y axis refer to the solid
  circles on the left panel (``all stars''). The surface number densities of the
  other star types are normalized to that of ``all stars'' at the outermost
  radial bin. Error bars are only drawn if they are larger than the symbol
  size. See discussion in \S\ \ref{s:rad_dist}.2.  
\label{f:rad_dists}}
\end{figure*}

After discarding the regions not dominated by cluster stars, we derived radial
distributions of the following regions on the CMD, all of which were found to
have less than 10\% contamination by field stars: {\it (i)\/} the lower MS,
meaning the MS below the turn-off of the 
field stars; {\it (ii)\/} the binary sequence associated with the lower MS;
{\it (iii)\/} the upper MS, meaning the MS below the turn-off area but above
the turn-off of the field stars; {\it (iv)\/} the binary sequence associated
with the upper MS; {\it (v)\/} the `lower half' of the MSTO region; {\it
  (vi)\/} the `upper half' of the MSTO region; {\it (vii)\/} the RGB and AGB
stars. Fig.\ \ref{f:regions} depicts these regions on the CMD with solid
lines, while the completeness-corrected radial distributions of stars in these  
regions are shown in Fig.\ \ref{f:rad_dists}. Only stars with completeness
fractions $\geq$\,50\% were used for this exercise.

Fig.\ \ref{f:rad_dists} shows several items of interest. First of all, a
comparison of the open squares with the filled squares in the left panel of
Fig.\ \ref{f:rad_dists} reveals that the lower MS is more heavily
contaminated by field stars than the upper MS. This could have been expected
given the presence of  
a wide turn-off near \B\ = 23.0 and \V\ = 22.5 (cf.\ Figs.\
\ref{f:fullCMDs}a,b), which is clearly due to field stars. A similar effect is
seen for the binary MS sequence (compare open circles with filled circles in
the right panel of Fig.\ \ref{f:rad_dists}). Second, a comparison of the
radial distribution of the upper MS stars with that of the bright RGB/AGB stars
(filled squares and filled triangles in the left panel of Fig.\
\ref{f:rad_dists}, respectively) shows that mass segregation is not
particularly strong in NGC~1846. There is a hint of the effect in the inner 
$\simeq$\,20\arcsec\ (equivalent to $\simeq$\,5 pc), 
where the surface number density of RGB/AGB stars continues to rise towards the
center at a slightly stronger rate than that of the upper MS stars. 
Last but not least, the right panel of Fig.\ \ref{f:rad_dists} shows that the
stars in the upper (brighter) half of the MSTO region {\it are\/} significantly more centrally 
concentrated in NGC~1846 than those in the lower (fainter) half of the MSTO. In fact,  
the stars in the upper half of the MSTO show a central concentration that is
similar to (or even stronger than) that of stars in the upper RGB/AGB. 
The very small difference in the luminosity and
mass (i.e., $\le 5$\%) of stars in the upper vs.\ the lower halves of the 
MSTO region strongly suggests that biases and 
mass segregation are {\it not} the cause of this difference.
We conclude that the upper and lower MSTO regions correspond to intrinsically
different populations which may well have undergone different amounts of
violent relaxation during their collapse and/or different dynamical evolution
effects. We discuss potential origins for this finding in \S\ \ref{s:disc}.    

\section{Isochrone Fitting} \label{s:isofits}  

We fit isochrones to the CMDs of NGC~1846
to determine the age, metallicity, \afe\
ratios, and their associated uncertainties.
Three sets of stellar models with predictions computed for the
ACS/WFC filter system were used:
Padova isochrones \citep{mari+08,gira+08}, Teramo isochrones
\citep[sometimes referred to as BaSTI isochrones;][]{piet+04,piet+06}, and
Dartmouth isochrones \citep{dott+08}.  

{\bf Padova isochrones:} We use the default models which involve scaled solar
abundance ratios (i.e., \afe\ = 0.0) and which include some degree of
convective overshooting \citep[see][]{gira+00}.
Using the web interface of the Padova
team\footnote{\tt http://stev.oapd.inaf.it/cgi-bin/cmd}, we construct a grid of
isochrones that covers the ages $0.3 \leq \tau [{\rm Gyr}] \leq 3.0$ (where
$\tau$ is the age) with a step of $\Delta \tau = 0.1$ Gyr and metallicities
$Z$ = 0.001, 0.002, 0.004, 0.006, 0.008, 0.01, 0.02, and 0.03. 

{\bf Teramo isochrones:}  We use that team's web
site\footnote{\tt http://albione.oa-teramo.inaf.it} to construct grids of
isochrones that cover the same ages and metallicities as for
the Padova models mentioned above. 
We use the Teramo isochrones with \afe\ = 0.0. 
To allow an assessment of the influence of different amounts of
convective overshooting, we construct  grids for both the
``canonical'' models which do not include overshooting and the
``non-canonical'' models which do.  

{\bf Dartmouth isochrones:} We use the full grid available from their web
site\footnote{\tt http://stellar.dartmouth.edu/$\sim$models/complete.html} which
covers the ages $0.25 \leq \tau [{\rm Gyr}] \leq 1.0$ with $\Delta\tau = 0.05$ Gyr and
$1.0 < \tau [{\rm Gyr}] \leq 5.0$ with $\Delta\tau = 0.25$ Gyr, metallicities
[Fe/H] = $-$2.5, $-$2.0, $-$1.5, $-$1.0, $-$0.5, 0.0, +0.3, and +0.5, and \afe\
= $-$0.2, 0.0, +0.2, +0.4, +0.6, and +0.8. 

\subsection{Fitting Method} \label{s:fitmeth}

Isochrone fitting is performed as follows. 
We first establish parameters that involve pairs of fiducial
points on the CMD that are: {\it (i)\/} relatively easy to measure or
determine from both the data and the isochrone tables, {\it (ii)\/} sensitive
to at least one population parameter such as age or metallicity, and {\it (iii)
  independent\/} on the distance and foreground reddening of the
cluster. This allows us to reduce the large number of free parameters,
which is always a concern when fitting isochrones to cluster CMDs.
After extensive experimentation, we select the following parameters: 
\begin{enumerate}
\item The difference in magnitude between the MSTO and the RGB bump, defined
  as $\Delta B_{\rm RGBB}^{\rm MSTO}$ and $\Delta V_{\rm RGBB}^{\rm
    MSTO}$ in the $B$ and $V$ filters, respectively. 
The MSTO is defined as the point where a polynomial fit
  to the stars (or the isochrone entries) near the turn-off is vertical in the
  CMD. For the location of the RGB bump on the CMD we simply calculate the
  mean magnitude and color of stars in a box centered on the RGB bump by eye. For the
  isochrones, we define the location of the RGB bump as the average of the 
  magnitudes and colors of isochrone RGB entries between the two
  masses at which the magnitudes and colors ``turn around'' in direction on
  the CMD with increasing stellar mass. 
\item The difference in color between the MSTO and the RGB bump, defined
  as $\Delta (B-I)_{\rm RGBB}^{\rm MSTO}$ and $\Delta (V-I)_{\rm RGBB}^{\rm MSTO}$. 
\item The slope of the RGB. This was evaluated using the (mean) color of the RGB
  stars at two fiducial magnitudes, namely at $m_{\rm RGBB} + 1$ and $m_{\rm
    RGBB} - 0.75$. The former magnitude was chosen to represent a point
  intermediate between the RGB bump and the lower end of the RGB; the latter
  magnitude was chosen to avoid issues related to confusing RGB with AGB stars
  on the CMD.   
  The mean colors of the RGB stars were derived from the CMD
  by means of a polynomial fit to the RGB star positions in the CMD. In case
  of the isochrones, the colors were derived by means of linear interpolation
  between isochrone table entries. 
\end{enumerate}

\begin{figure*}
\epsscale{0.65}
\plotone{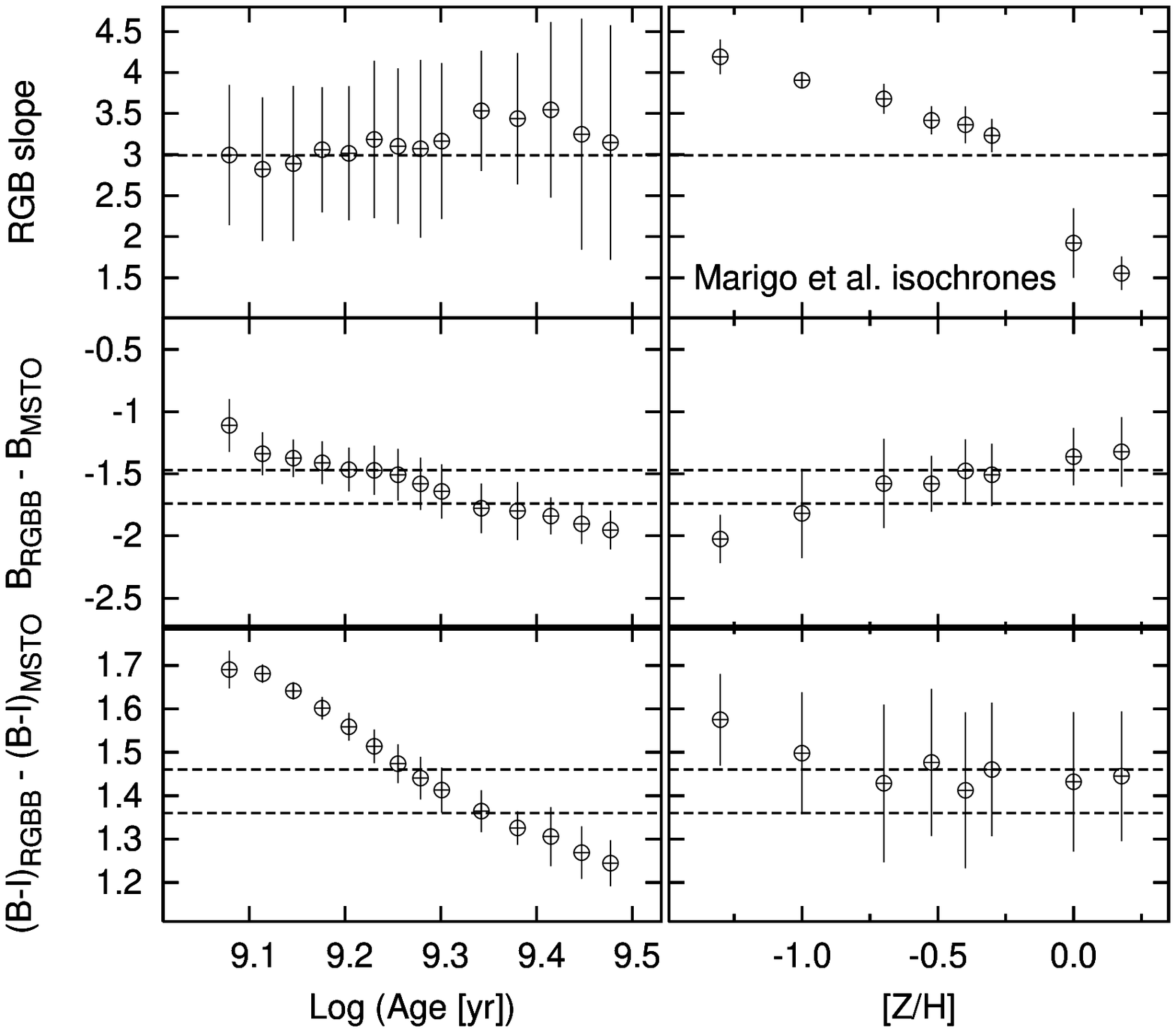}
\caption{Dependence of three distance-- and reddening-independent parameters
  involving fiducial points on the \B\ vs.\
  \BI\ CMD on age and metallicity for the Padova
  isochrones \citep{mari+08}. Top panels: Slope of the RGB as defined in \S\
  \ref{s:fitmeth}. Middle panels: $B_{\rm RGBB} - B_{\rm MSTO}$. Bottom
  panels: $(B-I)_{\rm RGBB} - (B-I)_{\rm MSTO}$. Error bars in the left and
  right panels reflect the variation of the parameter values among the
  isochrones with different metallicities and ages, respectively. The dashed
  lines in each panel represent the measurements of these parameters on the
  CMD. The upper and lower dashed lines in the middle and bottom panels refer
  to the upper and lower halves of the MSTO region, respectively. 
\label{f:agemetplot_girardi}}    
\end{figure*}

\begin{figure*}
\epsscale{0.65}
\plotone{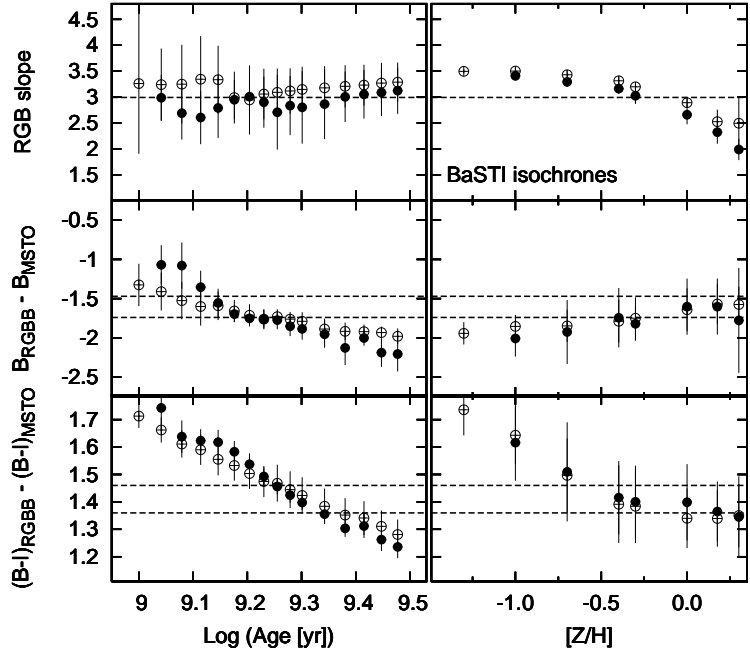}
\caption{Same as Fig.\ \ref{f:agemetplot_girardi}, but now for the Teramo
  isochrones \citep{piet+04}. The filled circles refer to the isochrones with
  convective overshooting, while the open circles refer to those without. 
\label{f:agemetplot_basti}}    
\end{figure*}

\begin{figure*}
\epsscale{0.65}
\plotone{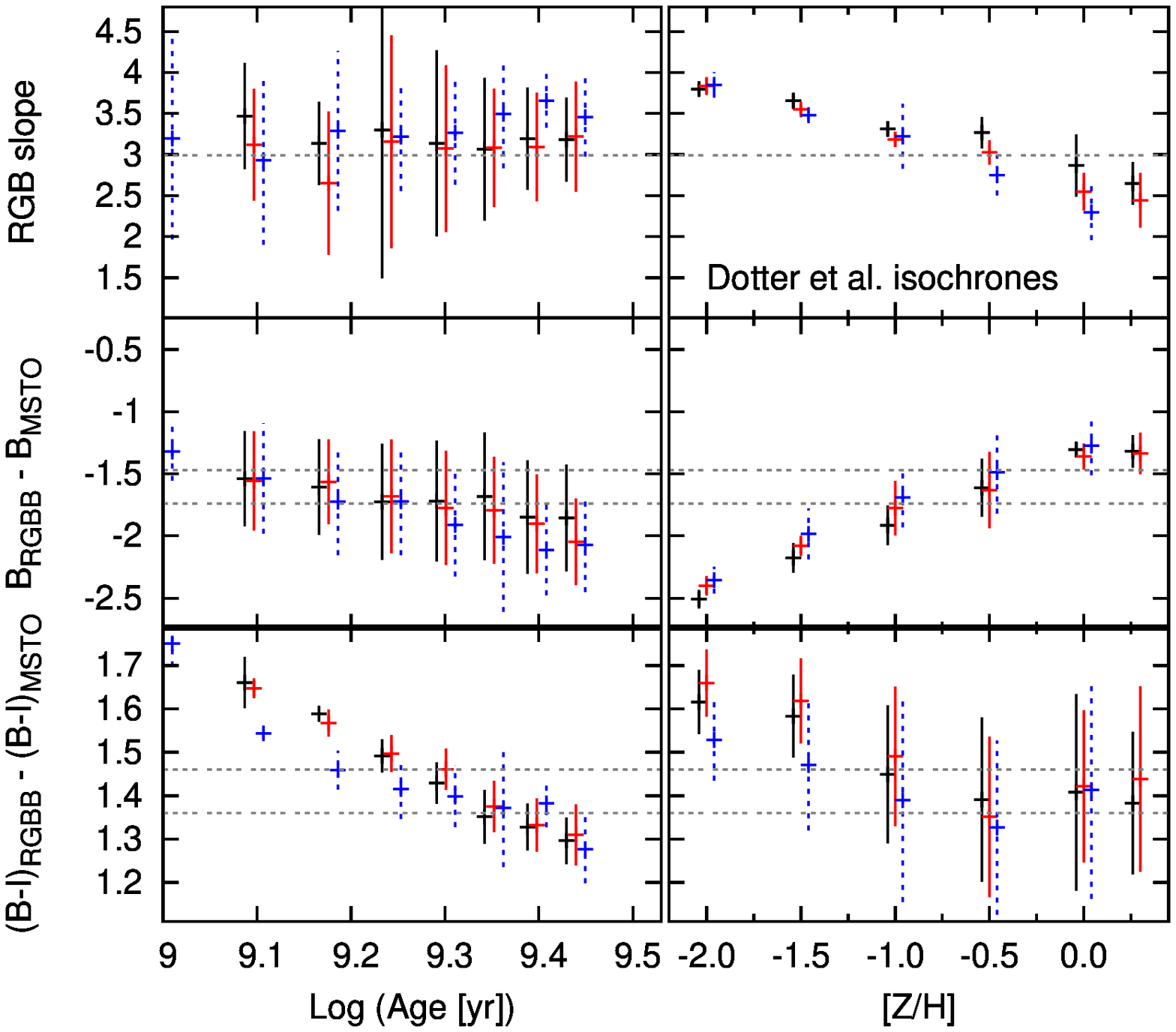}
\caption{Same as Fig.\ \ref{f:agemetplot_girardi}, but now for the Dartmouth
  isochrones \citep{dott+08}. The black symbols refer to the isochrones with
  \afe\ = 0.0, the red symbols do so for \afe\ = +0.2, and the blue symbols
  do so for \afe\ = +0.6. The three sets of symbols are plotted with small
  relative offsets in the $x$ direction to enhance clarity; the red symbols
  are plotted at the correct ages and [Fe/H] values. 
\label{f:agemetplot_dotter}}    
\end{figure*}

The sensitivity of these parameters to population parameters is
illustrated in Figs.\ \ref{f:agemetplot_girardi}\,--\,\ref{f:agemetplot_dotter} for
the age range 1\,--\,3 Gyr. These plots show the measurements of $\Delta
B_{\rm RGBB}^{\rm MSTO}$, $\Delta (B-I)_{\rm RGBB}^{\rm MSTO}$, and the RGB
slope as functions of age and metallicity (and \afe\ in case of the Dartmouth
isochrones) for isochrones in a given family, compared to the values measured
in the CMD of NGC\,1846. As might have been expected, the RGB slope
is highly sensitive to metallicity and almost independent of age in the age range
studied here. Figs.\
\ref{f:agemetplot_girardi}\,--\,\ref{f:agemetplot_dotter} also show that $\Delta
B_{\rm RGBB}^{\rm MSTO}$ is dependent on both age and \FeH, and that $\Delta
(B-I)_{\rm RGBB}^{\rm MSTO}$ is primarily dependent on age, although there
are also dependences on metallicity and \afe\ (see Fig.\
\ref{f:agemetplot_dotter} for the latter) among metallicities \FeH\ $\la
-1.0$ (the exact value of \FeH\ depending on the isochrone family used).  

We then select all isochrones (within each family) for which the values of
these three parameters lie within 2 $\sigma$ of the measurement uncertainty of
those parameters on the CMDs. 
This yielded 6\,--\,15 isochrones depending on the isochrone
family. For these isochrones, we then find the best-fit values for distance
$(m-M)_0$ and foreground reddening $A_V$ by means of a least squares fitting
program that runs through a grid of $(m-M)_0$ and $A_V$ values to compare the
magnitudes and colors of the MSTO and the RGB bump of
NGC\,1846 with those measured from the isochrones. The grid encompasses
$18.0 \leq (m-M)_0 \leq 19.0$ with a step of 0.01 mag 
and $0.0 \leq A_V \leq 0.5$ with a step of 0.01 mag. For the reddening we use
$A_{\rm F435W} = 1.351\, A_V$, $A_{\rm F555W} = 1.026\, A_V$, and $A_{\rm F814W} =
0.586\, A_V$. These values were derived using the ACS/WFC filter curves
available in the {\tt synphot} package of {\sc stsdas} and the reddening law of 
\citet{card+89} assuming $R_V = 3.1$. The resulting values of $(m-M)_0$ and
$A_V$ derived from the positions of the MSTO and RGB bump are averaged using
inverse variance weights. 

\begin{figure*}[htp]
\centerline{
\includegraphics[width=0.36\textwidth]{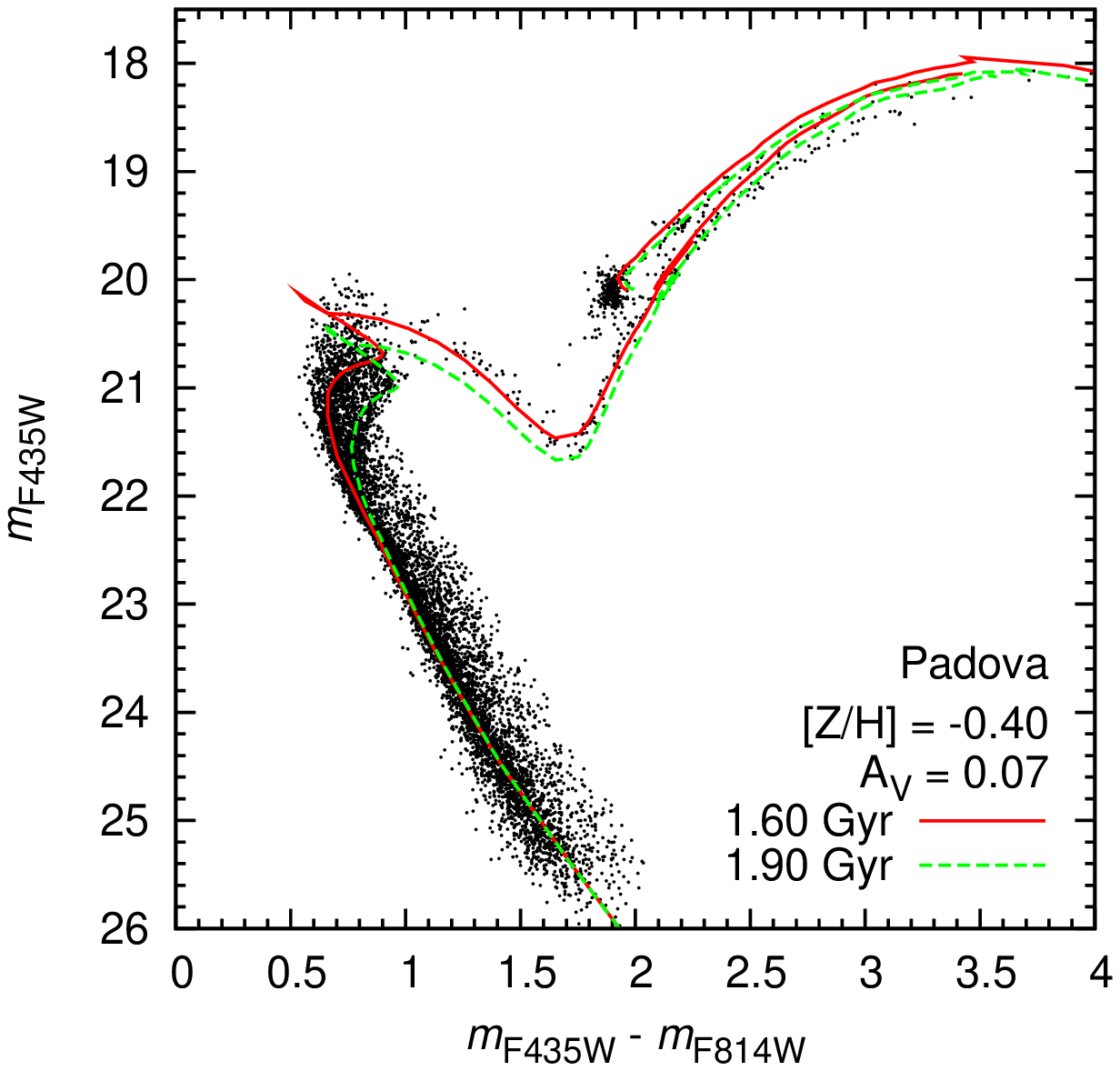}
\hspace*{0.1mm}
\includegraphics[width=0.36\textwidth]{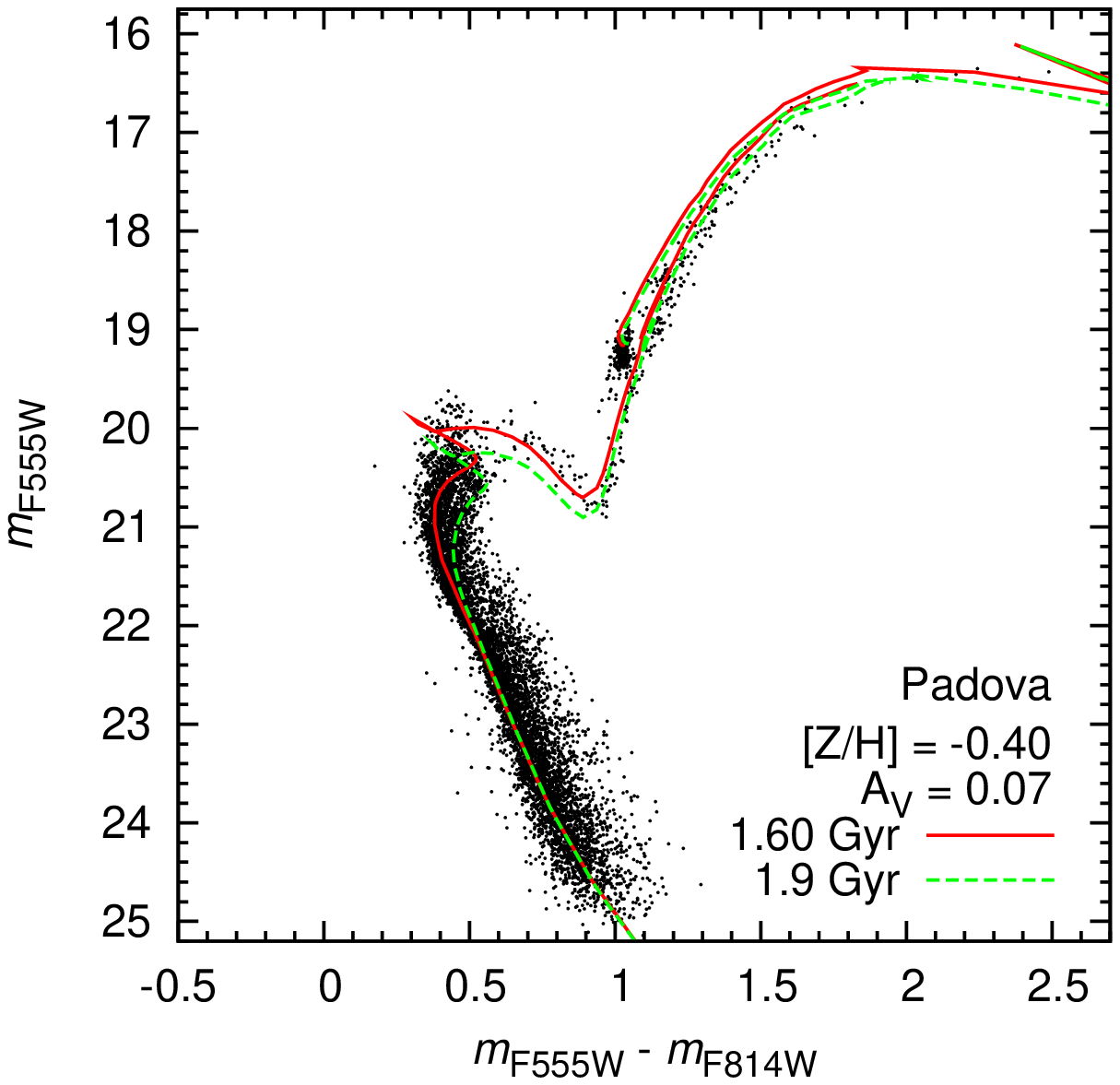}
}
\caption{Best-fit isochrones from the Padova family (cf.\
  Table~\ref{t:bestisotab}) superposed onto the CMDs
  shown in Fig.\ \ref{f:clusterCMDs}. The solid line is the best fit to the
  upper half of the MSTO region, while the dashed line is the best fit to the
  lower half of the MSTO region.
\label{f:bestiso_girardi}}   
\end{figure*}

\begin{figure*}[htp]
\centerline{
\includegraphics[width=0.36\textwidth]{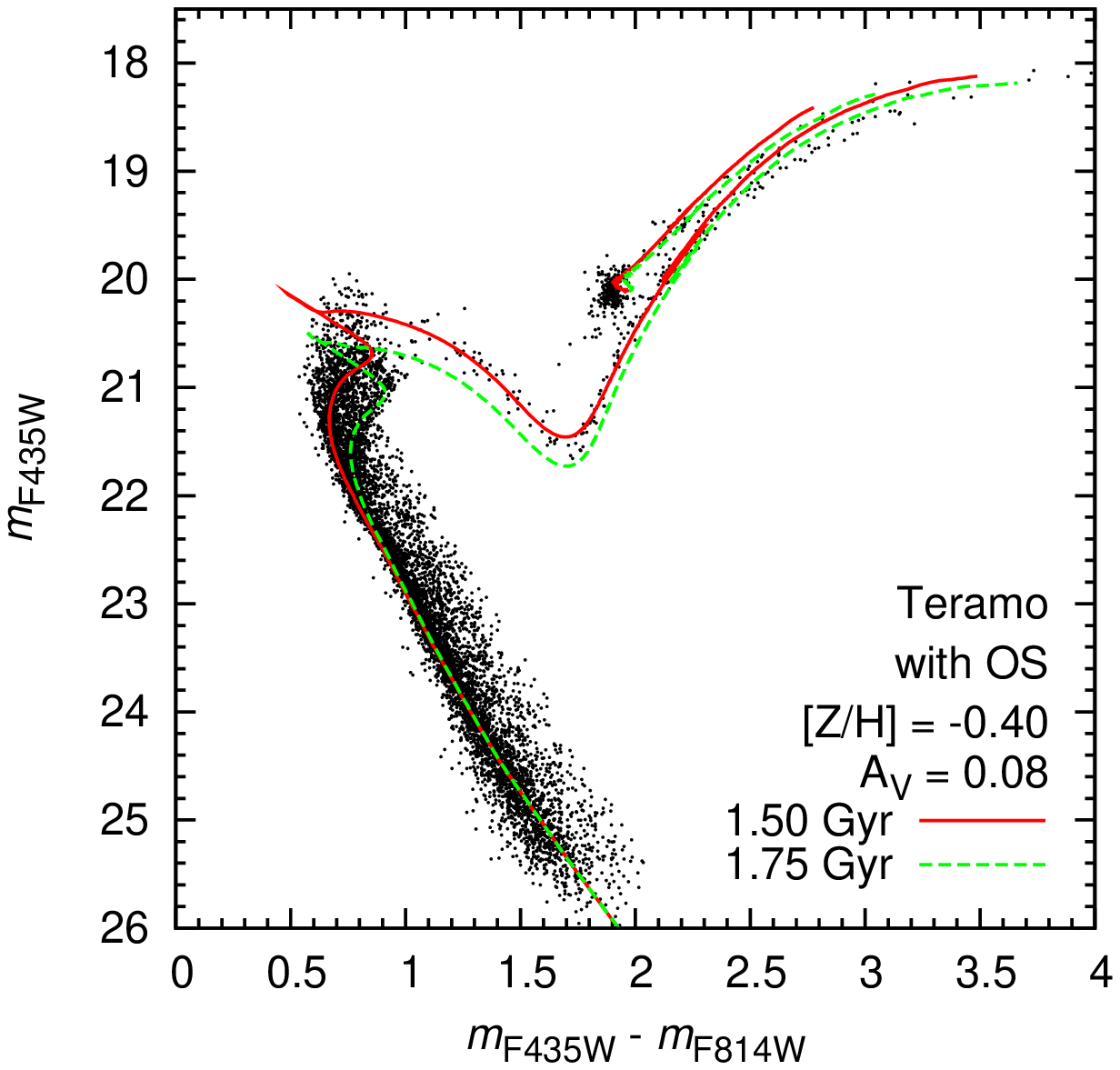}
\hspace*{0.1mm}
\includegraphics[width=0.36\textwidth]{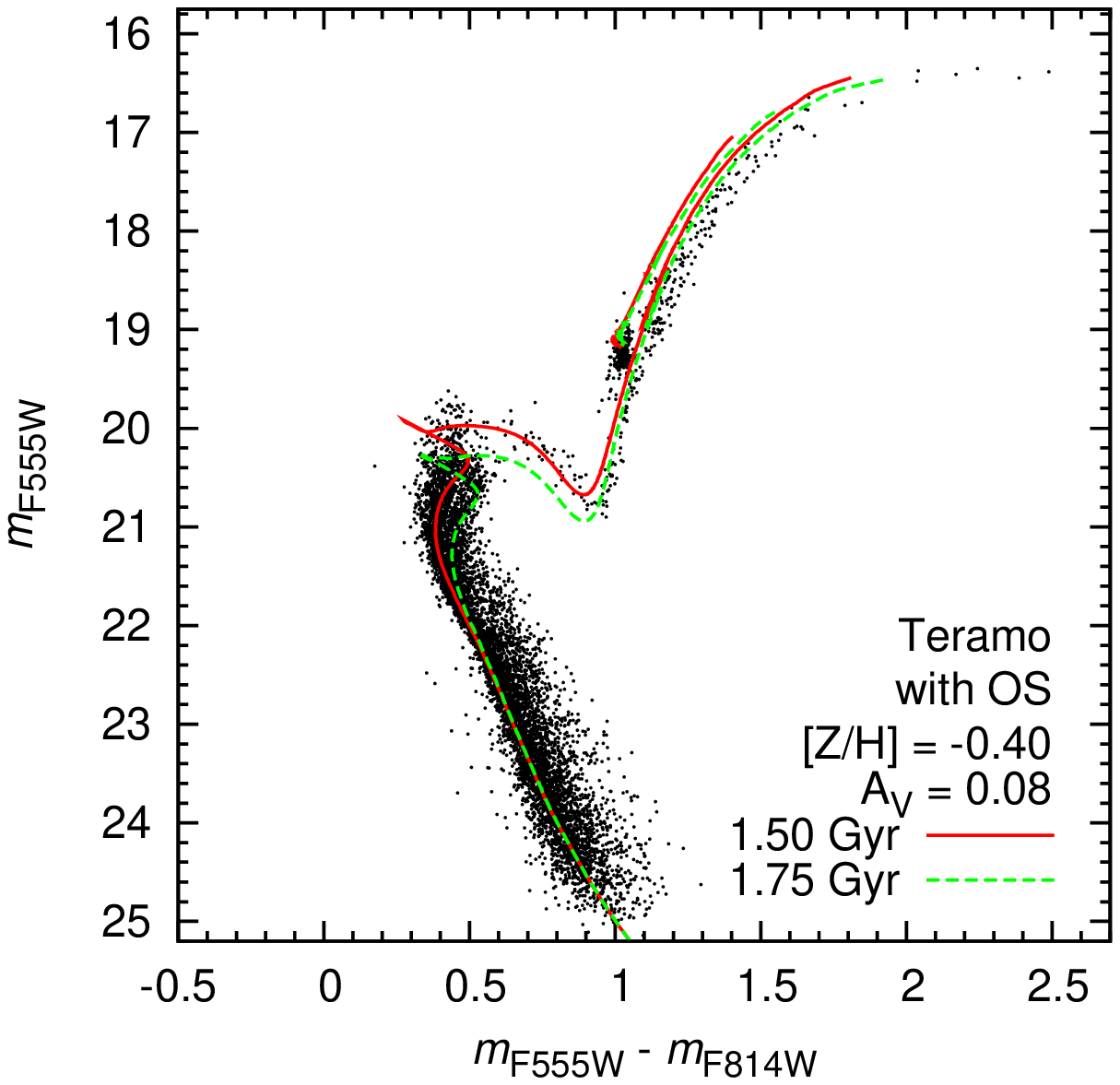}
}
\caption{Same as Fig.\ \ref{f:bestiso_girardi}, but now for the Teramo
  isochrones that include convective overshooting.
\label{f:bestiso_basti_OS}}   
\end{figure*}

\begin{figure*}[htp]
\centerline{
\includegraphics[width=0.36\textwidth]{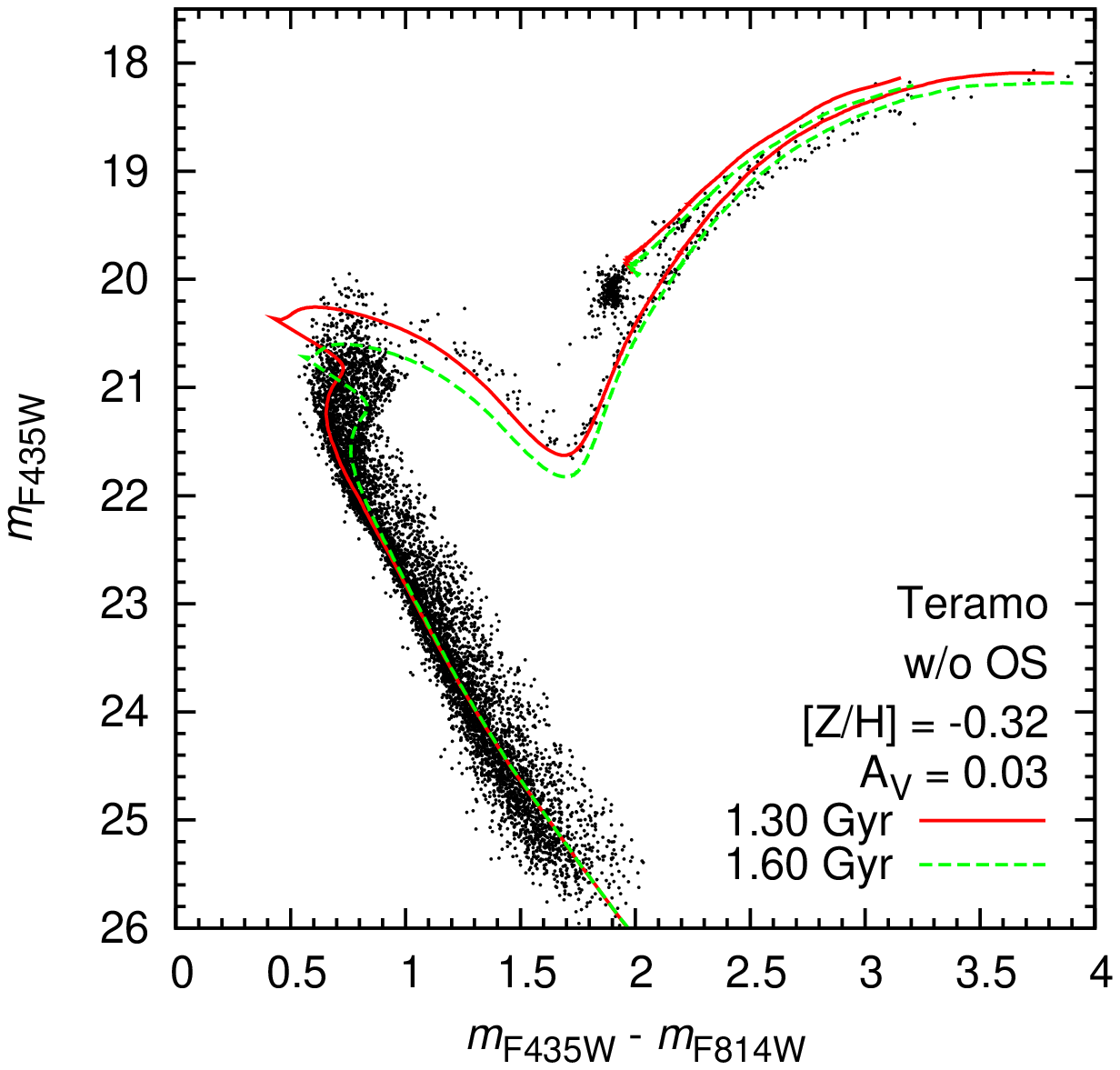}
\hspace*{0.1mm}
\includegraphics[width=0.36\textwidth]{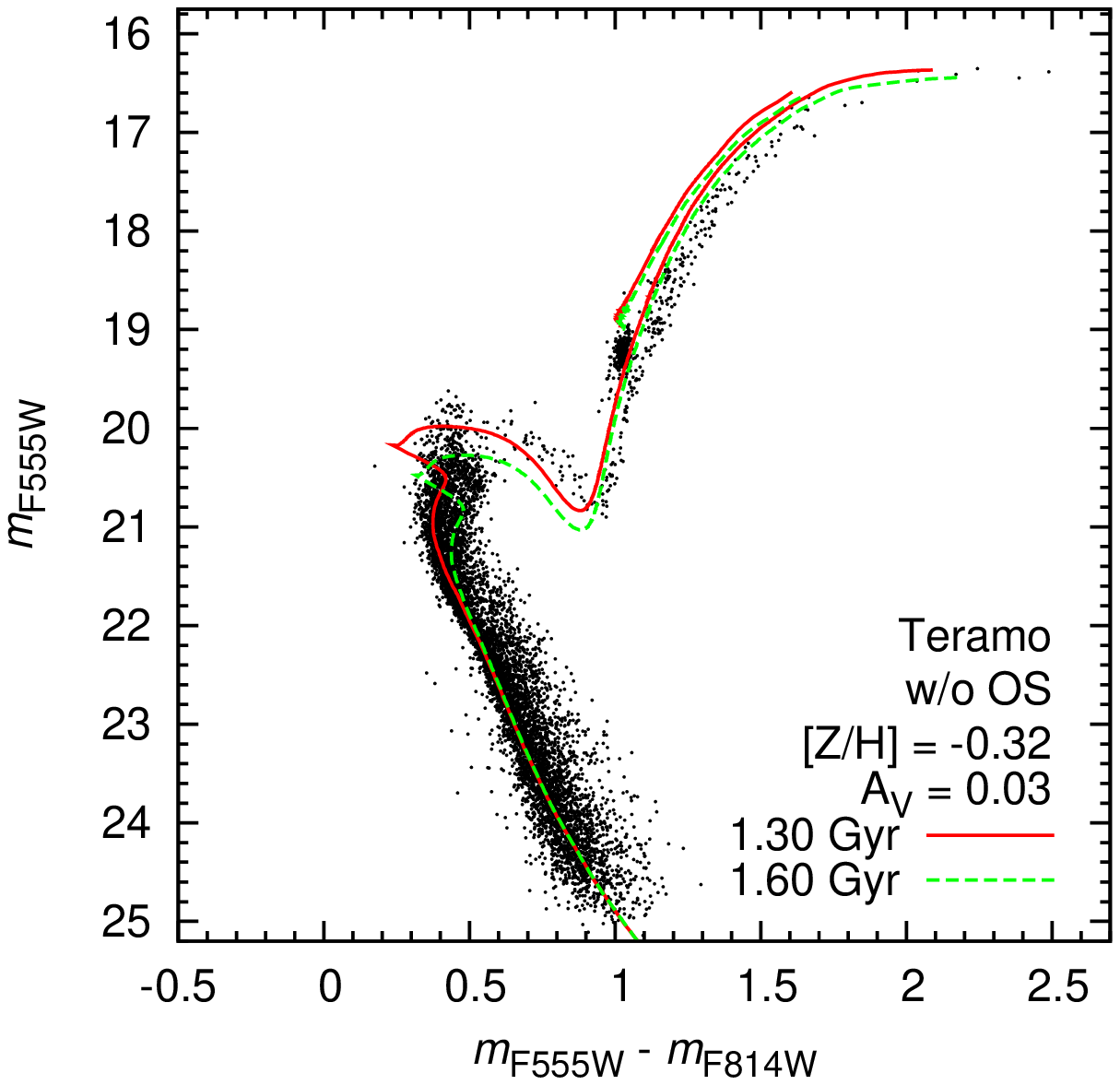}
}
\caption{Same as Fig.\ \ref{f:bestiso_girardi}, but now for the Teramo
  isochrones that {\it do not} include convective overshooting.
\label{f:bestiso_basti_noOS}}   
\end{figure*}

\begin{figure*}[ht]
\centerline{
\includegraphics[width=0.36\textwidth]{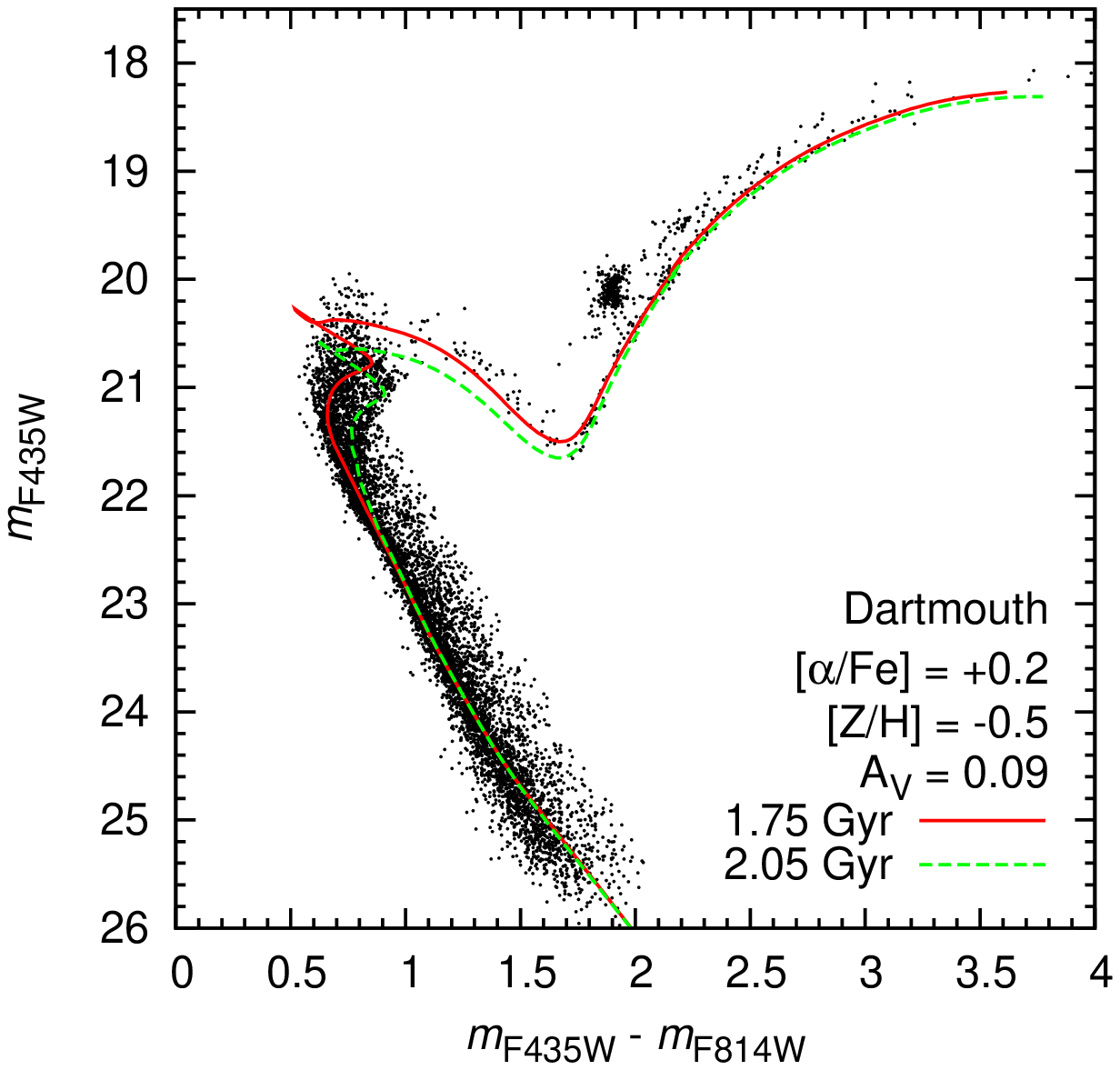}
\hspace*{0.1mm}
\includegraphics[width=0.36\textwidth]{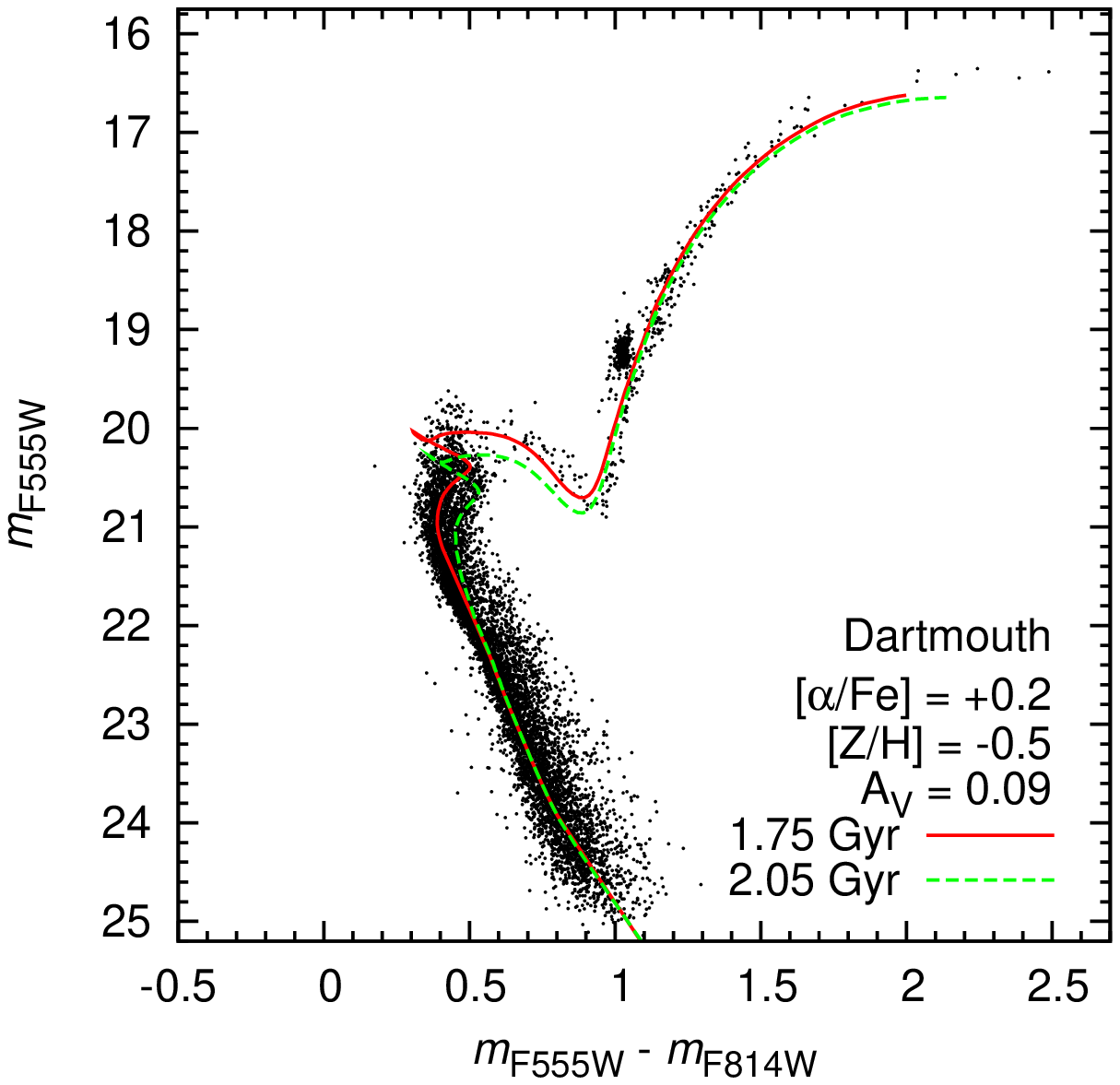}
}
\caption{Same as Fig.\ \ref{f:bestiso_girardi}, but now for the Dartmouth
  isochrones. 
\label{f:bestiso_dotter}}
\end{figure*}

Finally, the isochrones were overplotted onto the
CMDs for visual examination. This revealed that there was a small but
systematic offset in \FeH\ between the best-fit isochrones for \B\
vs.\ \BI\, and \V\ vs.\ \VI\ in the sense that
\FeH\ was always higher for the isochrone fit to \V\ vs.\ \VI\ 
than to \B\ vs.\ \BI. This effect was present for all isochrone
families used, although it was least significant for the Dartmouth
isochrones where the same \FeH\  yielded the best fit to both
CMDs; in the latter case the visual check merely showed a slight
mismatch of the RGB on the \V\ vs.\ \VI\ CMD (see Fig.\
\ref{f:bestiso_dotter}). 
We suggest that it is likely that this effect is related to 
the fact that Kurucz model spectra of cool RGB
stars contain more flux in the range $\sim$\,5000\,--\,6500 \AA\
(including much of the $V$ band) than empirical star spectra from the
\citet{pick98} library at the same stellar type, as recently reported by
\citet[][and private communication]{mara+08}. Model SEDs therefore have bluer
$V\!-\!I$ colors than observed for RGB stars, consistent with what
we see (see Figs.\ \ref{f:bestiso_girardi}\,--\,\ref{f:bestiso_dotter}).  
While a detailed investigation of the potential cause(s) of this effect is
beyond the scope of this paper, we note that the Dartmouth isochrones
used the {\sc Phoenix} model atmospheres \citep[e.g.,][]{haus+99} to
derive their $T_{\rm eff}$ -- color relations, whereas the Padova and
Teramo isochrones used the ATLAS9 models of R.\ L.\ Kurucz
\citep[e.g.,][]{caskur03}. The use of different model atmospheres can at 
least partly explain this effect since the {\sc Phoenix}
models include 550 million molecular transitions, while the ATLAS9
models only include a few molecular lines in a semi-empirical
manner. For example, the molecular band of MgH near 5175 \AA\ which is
in the $V$ band and strong in cool stars is not included in the ATLAS9
models.  
Because of this effect, we focus on the \B\ vs.\ \BI\ CMD in terms of
deriving population parameters and use the \V\ vs.\ \VI\ CMD mainly
for comparison purposes during the remaining analysis.   

In order to find the difference or spread in age that can explain the wide
morphology of the MSTO region of NGC\,1846, the fitting process mentioned
above was conducted separately for the MSTO's measured from the upper and
the lower halves of the MSTO region. During this process, we noticed that
the age step $\Delta \tau$ = 0.25 Gyr used for the grid of Dartmouth
isochrones is too coarse for this purpose. Hence, we created additional
Dartmouth isochrones covering a finer grid in age and metallicity around the
values found by the procedure mentioned above, using the web interface of
the Dartmouth team web site.

The best-fit isochrones and their population parameters (both for the upper and
the lower half of the MSTO region) are shown in Figs.\ 
\ref{f:bestiso_girardi}\,--\,\ref{f:bestiso_dotter}, superposed onto the
CMDs. While the best-fit isochrones of each model family generally match the various
stellar sequences well, there are some differences between the model
families. One such difference among the fits is seen in the MSTO region, where
the Teramo model without overshooting provides a poorer fit than the other three
models (which all do incorporate some level of overshooting). This is briefly
discussed in \S\ \ref{s:OS} below. Another, more significant difference
among the isochrone fits is seen on the upper RGB, where the Padova and Teramo
isochrones appear bluer than the observed stars whereas the Dartmouth isochrone is an
excellent fit to the full RGB. This difference is further discussed in
\S~\ref{s:aFe} below.  

\begin{table*}[htb]
\begin{center}
\scriptsize
\caption{Best-fit ages, metallicities, distances, reddening values,
  and [$\alpha$/Fe] ratio for NGC\,1846 using different isochrone
  families. See discussion in \S\ \ref{s:fitmeth} and \ref{s:syserrors}. 
 \label{t:bestisotab}}
\begin{tabular}{@{}lcccccccc@{}}
\multicolumn{3}{c}{~} \\ [-2.5ex]   
 \tableline \tableline
\multicolumn{3}{c}{~} \\ [-2.2ex]                                                
\multicolumn{1}{c}{Isochrone} & Age & Age & [Fe/H] & [$\alpha$/Fe] & $(m-M)_0$ & $A_V$ &
 \multicolumn{2}{c}{rms error\tablenotemark{a}}  \\ [0.2ex]
\multicolumn{1}{c}{Family} & (upper) & (lower) &    & & & & (upper) &
 (lower) \\ [0.5ex] \tableline  
\multicolumn{3}{c}{~} \\ [-2.5ex]              
Padova & $1.60 \pm 0.03$ & $1.90 \pm 0.03$ & $-0.40 \pm 0.05$ & (0.0) &
 $18.42 \pm 0.02$ & $0.07 \pm 0.02$ & 0.04 & 0.06 \\
Teramo w/ OS\tablenotemark{b}  & $1.50 \pm 0.03$ & $1.75 \pm 0.03$ &
 $-0.40 \pm 0.05$ & (0.0) & $18.50 \pm 0.02$ & $0.08 \pm 0.01$ & 0.07 & 0.03 \\
Teramo w/o OS\tablenotemark{c} & $1.30 \pm 0.05$ & $1.60 \pm 0.05$ &
 $-0.32 \pm 0.05$ & (0.0) & $18.43 \pm 0.02$ & $0.03 \pm 0.02$ & 0.03 & 0.02 \\ 
Dartmouth & $1.75 \pm 0.03$ & $2.05 \pm 0.03$ & $-0.50 \pm 0.05$ &
 $+0.2 \pm 0.1$ & $18.41 \pm 0.02$ & $0.09 \pm 0.02$ & 0.02 & 0.02 \\
Dartmouth\tablenotemark{d} & $1.95 \pm 0.03$ & $2.25 \pm 0.03$ & $-0.47 \pm 0.05$ &
 (0.0) & $18.40 \pm 0.02$ & $0.11 \pm 0.02$ & 0.03 & 0.03 \\ [0.3ex] \tableline
\multicolumn{3}{c}{~} \\ [-2.5ex]              
Adopted Values & $1.7 \pm 0.2$ & $2.0 \pm 0.2$ & $-0.5 \pm 0.1$ & $+0.2 \pm
 0.1$ & $18.45 \pm 0.05$ & $0.09 \pm 0.03$ \\ [0.5ex]
 \tableline
\end{tabular}
\tablenotetext{a}{RMS error of least-square fit to reference points on
the \B\ vs.\ \BI\ CMD, see discussion in \S\ \ref{s:fitmeth}}
\tablenotetext{b}{Teramo isochrones that include overshooting (OS)}
\tablenotetext{c}{Teramo isochrones that do not include overshooting}
\tablenotetext{d}{Dartmouth isochrones with [$\alpha$/Fe] = 0.0 only}
\end{center}
\end{table*}

The population parameters of the best-fit isochrones are listed in
Table~\ref{t:bestisotab} for each isochrone family. Table~\ref{t:bestisotab}
also lists the final adopted parameters for NGC~1846, including their
uncertainties that reflect both errors associated with the isochrone fitting
and systematic uncertainties related to the use of the different stellar models. 
The latter are discussed further in \S\ \ref{s:syserrors} below. 

\subsection{Influence of Overshooting Parameter} \label{s:OS}

Table~\ref{t:bestisotab} shows that the inclusion of convective overshooting
in isochrone models in the age range appropriate for GCs like NGC\,1846 has a
significant effect on the derived age and \FeH. In the case of the Teramo
isochrones, the inclusion of overshooting results in an increase in the
derived age of $\simeq$\,15\% and a decrease in the derived \FeH\ of
$\simeq$\,17\%. 

Focusing on the morphology of the MSTO region in the CMDs, a comparison of the
behavior of the best-fit Teramo isochrones that include overshooting with those
that do not (see Figs.\ \ref{f:bestiso_basti_OS} and 
\ref{f:bestiso_basti_noOS}) suggests that overshooting {\it is} occurring in
NGC\,1846. Hence we do not use the parameters of the best-fit Teramo isochrones
that do not include overshooting in evaluating the final adopted values of the
population parameters of NGC\,1846.  

\subsection{Influence of [$\alpha$/Fe] Abundance Ratio} \label{s:aFe}

\begin{figure*}
\centerline{
\includegraphics[bb=99 73 539 847,clip,angle=-90,width=0.7\textwidth]{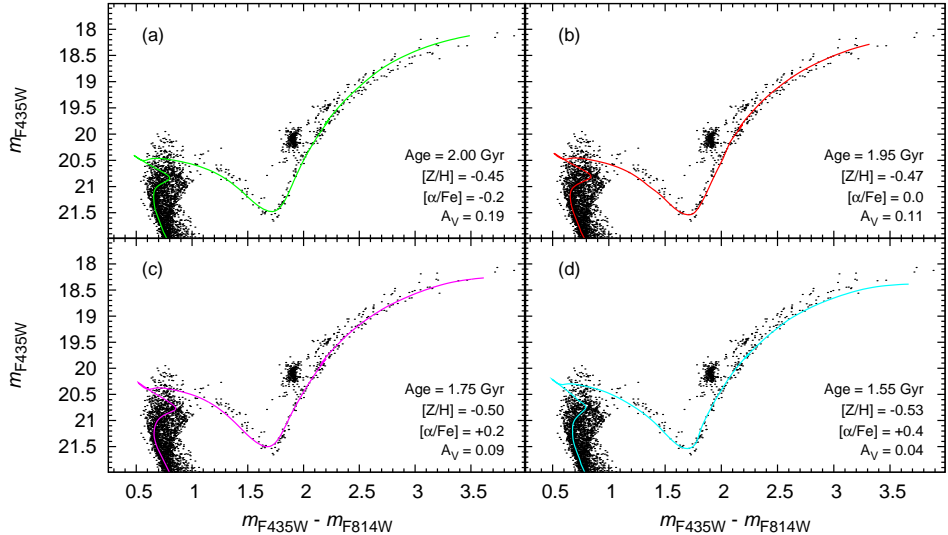}
}
\caption{Illustration of the effect of the \afe\ ratio on the
  isochrone morphology along the RGB as well as the derived ages and \FeH\ values. Panels
  (a) through (d) show the best-fit Dartmouth isochrones for \afe\ =
  $-$0.2, 0.0, +0.2, and +0.4 respectively. The legend lists the
  population properties of the isochrone shown in each panel. Note the
  significant effect of the \afe\ ratio on the slope and curvature of
  the RGB, as well on the resulting age and \FeH. 
\label{f:afeplot1}}
\end{figure*}

We assess the influence of non-solar \afe\ abundances on the derived age and
metallicity by comparing our CMD with Dartmouth ischrones for different values  
of \afe. The result is illustrated in Fig.\ \ref{f:afeplot1}, which shows the
best-fit Dartmouth isochrones for \afe\ = $-$0.2, 0.0, +0.2, and +0.4 superposed
onto the \B\ vs.\ \BI\ CMD of NGC\,1846. The fits were done to the upper half
of the MSTO region in this case.  Note that all four isochrones fit the MSTO and
RGB bump locations well, which is likely due to the fitting method we used (as
described above in \S\ \ref{s:isofits}). However, the detailed fits along
the RGB differ significantly from one \afe\ value to another. In
particular, there is a clear correlation between \afe\ and the slope or the
curvature of the RGB in the sense that larger \afe\  values give `flatter' slopes
and stronger curvature for the RGB. 

Furthermore, this comparison shows that larger values of
\afe\ result in younger fitted ages, and hence there is a degeneracy
between age and \afe\ if one doesn't take the detailed morphology of the
RGB into account. The amplitude of this 
effect is roughly 20\% in age for a difference in \afe\ of 0.2 dex (see Fig.\
\ref{f:afeplot1}). 
Finally, note that the effect of the chosen value of \afe\ to
the best-fit \FeH\ is small within the context of our fitting procedure:
$\Delta$\,\FeH/$\Delta$\,\afe~$\simeq$~$-$0.15. Note however that the effect
of non-solar \afe\ to determinations of [Fe/H] can be significantly larger for
analyses that aim to determine [Fe/H] values for more distant stellar systems
where only the upper end of the RGB is available for isochrone fitting. This
effect will be quantified in detail in a separate paper.  

As Fig.\ \ref{f:afeplot1} shows, the best fit to the full RGB is achieved
using the isochrone with \afe\ = +0.2. Since this fit is clearly better than
any isochrone that uses solar abundance ratios in any of the three families,
we adopt \afe\ = +0.2 for NGC\,1846.   

\subsection{Systematic Uncertainties in Derived Properties} \label{s:syserrors}  

We quantify systematic uncertainties in derived age, [Fe/H], 
distance, and reddening by comparing our best-fit results from
each set of isochrones with convective overshooting and \afe\ = 0.0,
as compiled in Table~\ref{t:bestisotab}.
These results give systematic uncertainties of:
$\pm$\,0.24 Gyr in age ($\simeq$\,15\%),
$\pm$\,0.05 dex in [Fe/H] ($\simeq$\,12\%), $\pm$\,0.05 mag in
$(m-M)_0$ ($\simeq$\,5\% in linear distance), and $\pm$\,0.03 mag in
$A_V$ ($\simeq$\,30\%). 
We suggest that these values represent typical
systematic uncertainties associated with the determination of 
population parameters of intermediate-age star clusters from \B\ vs.\ 
\BI\ CMD fitting by isochrones of any given stellar model. 

Finally, we note that the properties of the best-fit isochrones are
consistent with an estimate of the mass of AGB stars in NGC 1846
as derived from their pulsational properties:\ \citet{lebwoo07} 
estimated ${\cal{M}}_{\rm AGB} \simeq 1.8\; \mbox{M}_{\odot}$ with an
uncertainty of $\simeq 0.2\; \mbox{M}_{\odot}$ (Lebzelter, private
communication). The 
formal AGB star masses for the best-fit isochrones in Table~\ref{t:bestisotab}
all fall between 1.70 and 1.77 M$_{\odot}$, consistent with the independent
estimate of Lebzelter \& Wood.

\section{Discussion} \label{s:disc}

\subsection{Morphology of the MSTO Region} \label{s:MSTOmorph}

As already mentioned in \S\S\ \ref{s:CMDs} and \ref{s:rad_dist}.2, the MSTO 
region is broader than can be explained by photometric 
errors, and the upper (brighter) 
MSTO region has a significantly more centrally 
concentrated radial distribution than the lower one. This finding 
strongly suggests the presence of multiple (at least two) stellar populations 
in NGC 1846. To constrain this picture further in terms of whether or
not two distinct populations are sufficient (or
  necessary) to produce the observed CMD, we conduct simulations of  
a synthetic cluster with the properties implied by the isochrone
fitting in \S\ \ref{s:isofits}. For this purpose we define the null
hypothesis such that the upper and lower halves of the MSTO region
represent two distinct simple stellar populations with properties
given in Table~\ref{t:bestisotab} for the best-fit Dartmouth isochrones. 

We simulate a simple stellar population (SSP) with a given age and
chemical composition by populating an isochrone with stars randomly drawn
from a Salpeter mass function between 0.1 M$_{\odot}$ and the RGB-tip
mass, which is at 1.62 M$_{\odot}$ for the best-fit Dartmouth isochrone
(see 4th entry of Table~1). The total number of stars is normalized to the
observed number of stars brighter than the 50\% completeness limit. To a
fraction of this sample of stars we add unresolved secondary binary
components that are drawn from the same mass function, hence assuming a
flat primary-to-secondary mass ratio distribution for binary stars. We note
that, although we are primarily investigating the morphology of the
MSTO region, we fully model the stellar evolution until the onset of
He-shell burning and also add secondary binary components that were
evolved past the tip of the RGB. Their evolution is modeled using the
He-core burning evolutionary tracks assuming an average mass-loss of 0.15
M$_{\odot}$ along the RGB and a stellar mass dispersion on the zero-age
horizontal branch of 0.06 M$_{\odot}$ \citep[see][for details]{Lee91,
Dotter08}. We use the width of the upper main sequence, i.e. the part
brighter than the the turn-off of the background stellar population, to
fix the binary star fraction at 15\%. We estimate the internal systematic
uncertainty in binary fraction to $\pm5\%$ and defer a more detailed
discussion of the parameter degeneracies, in particular binarity vs. mass
fractions of stellar generations (i.e.,~the star formation history of
the cluster), to a future paper. For the purposes of this work the 
results don't change significantly within $\sim\!10\%$ of the binary
fraction. Finally, we add photometric errors that are modeled using the
distribution of photometric uncertainties of our observations to the
sample of artificial stars.

We compare the observed MSTO morphology of NGC~1846 with those of the simulated 
CMDs by constructing a parallelogram in the CMD with the following
characteristics: {\it (i)\/} One axis is approximately parallel to
both isochrones, {\it (ii)\/} the other axis is approximately perpendicular to
both isochrones, and {\it (iii)\/} it is located in a region of the MSTO
where the split between the two isochrones is relatively evident. The
comparison between data and simulations is then done by transforming
the (\BI, \B) coordinates of the stars in the CMD into the reference
coordinate frame defined by the two axes of the parallelogram, and
then considering the distributions of reference coordinates of the
stars in the direction perpendicular to the isochrones. 

\begin{figure*}[thb]
\centerline{
\includegraphics[bb=99 58 490 733,clip,angle=-90,width=0.75\textwidth]{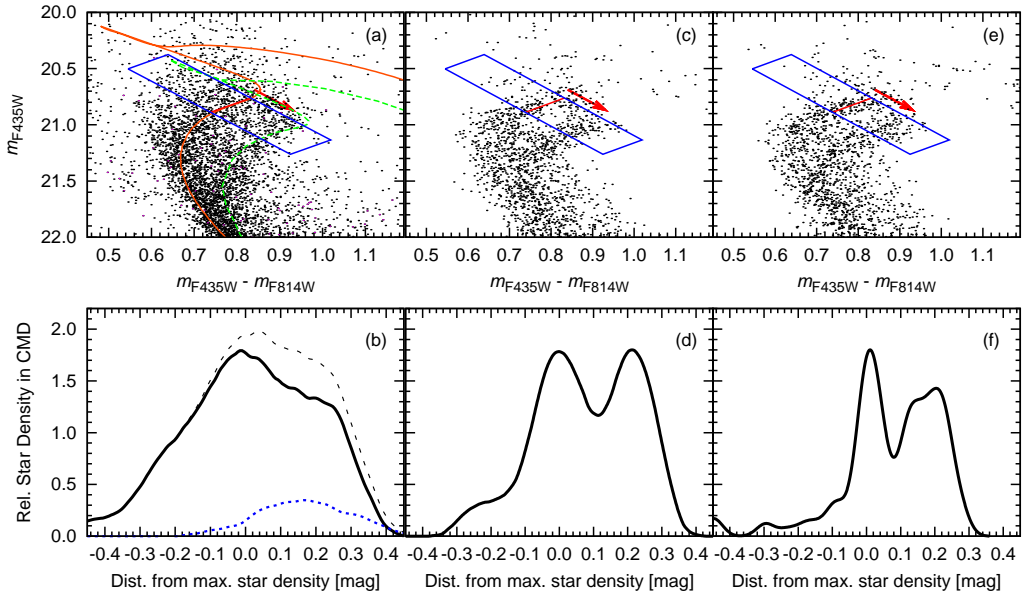}
}
\caption{ 
Panel (a): Enlargement of the CMD shown in Fig.\ \ref{f:fullCMDs}a focusing on the
MSTO region. The orange and green curves represent the best-fit 
Dartmouth isochrones to the upper and lower halves of the MSTO region, 
as shown in Fig.\ \ref{f:bestiso_dotter}. Panel (b): Distribution of
stars in the parallelogram shown 
in the top panel, derived using a non-parametric density estimator (see text for
details). The black dashed line represents all stars, the blue dashed line
represents stars in the background area (scaled to the full area of the ACS
image), and the solid line represents ``all stars minus background''. The red
arrow in panel (a) indicates the positive direction of the 
X axis of panel (b), while the red line indicates its zero point. 
Panels (c) and (d): Same as panels (a) and (b), respectively, but for the
simulation of two SSPs including binary stars as described in \S\ 
\ref{s:MSTOmorph}, with 50\% of the stars belonging to the 1.75 Gyr old SSP. 
Panels (e) and (f): Same as panels (c) and (d), respectively, but now
for the case where 60\% of the stars belong to the 1.75 Gyr old SSP. 
\label{f:acrossMSTOplot}}
\end{figure*}

The procedure mentioned above is illustrated in
Fig.~\ref{f:acrossMSTOplot}. The top panels (a, c, and e) 
show the CMDs of the data and two simulations with relevant fractions
of stars in the two distinct SSPs (in black dots) along with the
isochrones (in orange and green) and the 
parallelogram mentioned above (in blue, with the reference axis along
the isochrones in red). The bottom panels (b, d, and f) show the
distributions of the stars' coordinates in the direction
perpendicular to the isochrones for the CMDs shown in the
corresponding top panels. To avoid issues related to the use of finite
histogram bin sizes in the presence of relatively few data points, the
coordinate distributions were calculated by means of the
non-parametric Epanechnikov-kernel probability density function 
\citep{silv86} for all objects in the parallelograms. In the case of
the observed CMD (i.e., panels a and b), this was done both for stars
within a radius of 42\farcs5 from the cluster center and for the
``background region'' for which the CMD 
was shown in Fig.~\ref{f:fullCMDs}e. The area of the probability
density function of the background region was then scaled to that of
the inner 42\farcs5 radius by area on the ACS image, and the true probability
density function of the inner 42\farcs5 radius was derived by statistical
subtraction of the background region (see panel b). 

As panel (b) of Fig.~\ref{f:acrossMSTOplot} shows, the
distribution of observed stars in the parallelogram peaks at the
location of the ``younger'' isochrone and then declines more or less 
uniformly towards higher coordinate values, i.e., older ages. A
comparison with panels (d) and (f) shows that the distribution of the
observed stars is clearly more uniform than those of the two-SSP
simulations. Two-sample Kolmogorov-Smirnov (K-S) tests confirm this
visual impression: The $p$ values of the K-S tests to compare the
distribution of the simulated CMDs of two SSPs with various mass
fractions of the younger population against the data never exceed 
11\% (see Table~\ref{t:KStest}). It thus seems fair to conclude from these
data that the observed CMD of NGC\,1846 can be better explained by a
population with a {\it distribution of ages\/} rather than by two discrete
SSPs as suggested by \citet{mack+08} and \citet{milo+08b}.  

\begin{table}[hbt]
\begin{center}
\caption{Results of two-sample Kolmogorov-Smirnov tests regarding the morphology of the
  MSTO region. Null hypothesis is that the data of NGC~1846 and the two-SSP
  simulations listed in the table are drawn from the same parent
  distribution. See discussion in  \S\ \ref{s:MSTOmorph}.    
 \label{t:KStest}}
\begin{tabular}{@{}cc@{~~}cc@{}}
\multicolumn{2}{c}{~} \\ [-2.5ex]   
 \tableline \tableline
\multicolumn{2}{c}{~} \\ [-2.2ex]                                                
\multicolumn{1}{c}{$f_{1.75 {\it Gyr}}$\tablenotemark{a}} & $p$ value &
\multicolumn{1}{c}{$f_{1.75 {\it Gyr}}$\tablenotemark{a}} & $p$ value  \\ [0.5ex] \tableline
\multicolumn{2}{c}{~} \\ [-2.5ex]              
0.25 & 0.01 & 0.55 & 0.11 \\
0.30 & 0.01 & 0.60 & 0.11 \\
0.35 & 0.06 & 0.65 & 0.04 \\
0.40 & 0.10 & 0.70 & 0.03 \\
0.45 & 0.11 & 0.75 & 0.01 \\
0.50 & 0.11 &  & \\ [0.5ex]  \tableline
\end{tabular}
\tablenotetext{a}{Mass fraction of stars in the (younger) SSP (age\,=\,1.75
  Gyr) in the two-SSP simulation, see \S\ \ref{s:MSTOmorph}}
\end{center}
\end{table}

\subsection{On The Formation of NGC 1846} \label{s:nature}

The results described above naturally raise the question: How did
an intermediate-age star cluster like NGC~1846 with its wide MSTO form?
Is its formation related in some way to the ancient globular clusters
in our own Galaxy for which evidence suggests the presence of multiple stellar 
populations? 
Based on the results of the experiments described above, we assume that the 
wide MSTO in NGC~1846 is due to a {\it spread} in the ages of stars
found within this cluster.
There are two general situations which can lead to the presence of stars
with a wide range of ages within a single cluster: one involving stars
which formed originally outside of the cluster and then were accreted in
some way, and another where stars formed over an extended period
within the cluster itself.

Three specific scenarios have been proposed for an external origin of
stars with unusual CMDs: {\it (i)\/} the merger of two (or more) star
clusters, {\it (ii)\/} the merger of a star cluster with a giant
molecular cloud (hereafter GMC), and {\it (iii)\/} the entrapment of pre-existing
field stars during cluster formation.
As to the possibility that NGC~1846 resulted from the merger
of two star clusters, the LMC is known to host tens of young
(10$^7$\,--\,10$^8$ yr old) candidate double star clusters
\citep[e.g.,][]{bhahat88}, many of which have been shown to be physically
linked objects \citep[e.g.,][]{leon+99}. 
On the other hand, most formation scenarios
\citep{fujkum97,thei02} suggest that binary clusters form
within $\approx10$~Myr, much faster than the age range seen within
NGC~1846. 
However, it has been suggested  \citep{macbro07}
that star clusters which form in groups \citep[star cluster groups or 
SCGs, see][]{leon+99} inside GMCs can have ages separated by 
a longer period of time. 
This idea seems consistent with the finding of \citet{efrelm98} that 
the average age difference between 		
pairs of star clusters in the LMC increases systematically with their
(deprojected) angular separations, suggesting that the star formation process is 
hierarchical in space and time, i.e., that larger star-forming regions form
stars over longer periods than small regions. However, the very small 
star-to-star variation in [Fe/H] in NGC~1846 allowed by the narrowness
of its RGB seems difficult 
to account for in this scenario: Fig.\ 1 of \citet{efrelm98} shows that
an average age difference between star cluster pairs of $\simeq$\,200 Myr
corresponds to angular separations $\ga$\,0\fdg2, or $\ga$\,150 pc
at the distance of the LMC. This size is larger than that of any GMC found in
the Galactic surveys reviewed in \citet{efrelm98}. 
Coupled with our finding above that the main sequence turnoff region
in NGC~1846 does not appear bimodal as much as it appears continuous,
a binary cluster merger origin for this cluster seems unlikely.

Recent simulations of mergers of star clusters with GMCs showed that
such mergers can form a second-generation of stars that is (at least
initially) more centrally concentrated than the first generation of stars
\citep{bekmac09}, which is consistent with what we find in NGC~1846. 
However, to reproduce the
observed features in intermediate-age star clusters with broad MSTO
regions, this scenario seems to require rather strongly constrained
ranges of GMC parameters such as their spatial distribution, mass
function, dynamics, and chemical composition at the time the mergers would 
have occurred. For example, all currently known star clusters in the
LMC with broad MSTO regions exhibit a range in age of
$\sim 1 - 3 \times 10^8$ yr \citep[cf.][]{mack+08,milo+08b}. Larger
age differences have so far not been 
found. This would require significant fine-tuning of the range of
distances between the GMCs and the initial star clusters to be merged and/or
their relative velocities, 
whereas the current locations of the star clusters with broad MSTO
regions cover a rather wide range within the LMC. Another constraint
implied by this scenario is that the GMCs would all have to have
the same [Fe/H] as the star clusters with which they merged, given the
very small range in [Fe/H] allowed by the narrowness of the RGB in
star clusters like NGC 1846. 

Other simulations suggest that during formation, star clusters can
trap pre-existing field stars, leading to the presence of stars with a
range of ages \citep{fell+06,pflkro07}. 
This scenario also makes predictions which are consistent with
observations: {\it (i)\/} the radial surface density of the trapped (older) 
stars will be flatter than that of stars formed in the cluster itself, and 
{\it (ii)\/} the absence of clear peaks in the distribution of stars across the
MSTO region also seem compatible with this scenario. 
The biggest point against this scenario is that the simulations of 
\citet{fell+06} indicate that a star cluster
with an initial mass of $\simeq$ 10$^6$ M$_{\odot}$ would only be able to trap
up to a few percent of its initial mass. This is inconsistent
with the 30\,--\,40\% of the stars within NGC~1846 which appear
to belong to an older subpopulation.
The apparent homogeneity in [Fe/H] of all stars in
NGC~1846 also seems hard to explain within the field star trapping scenario.

We find that a situation where all stars formed in situ is in good agreement
with several different findings for NGC~1846 noted in this paper.
In particular, we suggest a scenario where gas lost by means of low-velocity
winds of stars with masses lower than those producing supernovae (SN)
of type II\footnote{NGC~1846 has an estimated mass
of $\approx2\times10^5~M_{\odot}$, based on an age of 1.7~Gyr, [Fe/H]$=-0.50$,
and either the \citet{mara05} or \citet{bc03} models assuming a 
Kroupa and Chabrier IMF, respectively.
This mass is approximately an order of magnitude lower than that necessary to
retain the chemically enriched gas expelled during the earliest stages of the
first collapse \citep[e.g.,][]{recdan05,basgoo06}.} accumulates in the central regions
of the star cluster and forms a second generation of stars on timescales 
similar to the observed age difference between the upper and lower 
end of the MSTO in NGC~1846. 
The stellar agents responsible for such
self-enrichment may be {\it (a)\/} slow winds of fast-rotating O-- and
B-type stars, which occur at ages of 1\,--\,10 Myr \citep{decr+07} or
{\it (b)\/} low-velocity winds of intermediate-mass
($\simeq$\,3\,--\,7 M$_{\odot}$) stars in the AGB phase, which have
ages of 100\,--\,300 Myr \citep{vent+01,vent+02,vendan08}.  Note that
the time scale for intermediate-mass AGB stars to form in the first
population is similar to the actually observed age difference
between the upper and lower end of the MSTO in NGC~1846. 
\citet{derc+08} recently performed hydrodynamical
and $N$-body simulations to study the dynamical evolution of star clusters in
the context of such a self-enrichment scenario. Interestingly, the
D'Ercole et al.\ results show that clusters with masses, ages, and
structural parameters similar to those of NGC~1846 are expected to
show a second-generation population that is significantly more
centrally concentrated than the first-generation stars (see in
particular their Figure 18), just as we see in NGC~1846. This would
also be consistent with the observed effect that the stars in the
upper part of the MSTO are more centrally concentrated than stars in
the RGB and AGB (see Fig.\ \ref{f:rad_dists}), since the latter are
expected to contain stars from both populations.  
The slow winds from either fast-rotating massive stars or intermediate-mass
AGB stars would also lead to chemical enrichment of light elements in
the subsequent generation of stars due to products from the CNO cycle.  
If this scenario is correct, then one would expect to see significant
and correlated variations in the light 
element abundances (e.g., C, N, O, F, Na, Mg) among the stars in NGC 1846, 
perhaps similarly to those found in several Galactic GCs.
Measuring the chemical composition of RGB stars in NGC~1846 and other
LMC clusters with wide MSTOs \citep[e.g.,][]{bert+03,milo+08b} is
feasible with current spectrographs on 8-10m-class telescopes and
should provide additional relevant evidence to help decipher the most
likely formation scenario. 

\section{Summary and Conclusions} \label{s:conc}

We have used deep {\it BVI} photometry from {\it HST/ACS} images to construct
CMDs of the intermediate-age star cluster NGC~1846 in the LMC.  We have used the
{\it ePSF\/} fitting technique developed by J.\ Anderson \citep{andkin06},
which returns high-accuracy photometry of cluster stars extending some 5
magnitudes below the main sequence turnoff for this cluster. 

The CMD for NGC~1846 shows a number of interesting features (many of
which have been noted previously):
{\it (i)\/} a very narrow RGB; 
{\it (ii)\/} an MSTO 
region that is clearly broader than the (fainter) single-star main sequence; 
{\it (iii)\/} an obvious sequence of unresolved binary stars, somewhat
brighter and redder than the single star main sequence.
We have performed a number of new experiments, including checking
the effects of background subtraction, completeness corrections, and
binary star evolution, which confirm that the cluster has a
statistically significant broad MSTO region.

We fit isochrones from the Padova, Teramo, and Dartmouth groups in order
to determine the best-fit population parameters for NGC~1846. We
perform two fits with each set of isochrones: One for the upper half
of the MSTO region and one for the lower half. 
All three sets give a reasonably good fit to the CMD, although there
are differences between the observations and predictions in the shape
of the RGB for scaled-solar abundance ratios. 
The overall best fit to the entire CMD is achieved using the Dartmouth
isochrones. The adopted parameters for NGC~1846 are:
Distance $(m-M)_0$ = 18.45, reddening $A_V = 0.09$, ages of 1.7 Gyr
(upper half of MSTO) and 2.0 Gyr (lower half), metallicity [Fe/H] =
$-$0.50, and a non-solar abundance ratio [$\alpha$/Fe] = +0.2.  

We use the results from the isochrone fitting to quantify typical systematic
errors of fitted population parameters for intermediate-age star
clusters like NGC~1846 introduced by using any one family of
isochrones, the inclusion or exclusion of convective overshooting, and
the assumption of solar abundance ratios. These systematic errors are
typically of order 15\,--\,30\% for any given output parameter mentioned above. 

We find for the first time that the
radial distributions of stars of different masses/ages vary 
with statistical significance within NGC~1846, with stars in the upper
MSTO (i.e., the younger stellar generation) being more centrally
concentrated than stars in any other  
area of the CMD, including more massive RGB and AGB stars. 
Since this cannot be due to dynamical evolution of a simple stellar
population, we conclude that the upper and lower MSTO regions
correspond to intrinsically different populations which have undergone
different amounts of violent relaxation during their collapse. 

We have tested the role played by binary evolution on stars in the MSTO
region of the CMD via Monte-Carlo simulations of multiple stellar generations.
Our multi-SSP models include late stellar evolutionary stages past the
He-core flash, a realistic treatment of photometric uncertainties,
and a flat distribution of primary-to-secondary stellar mass ratios for binary stars.  
A quantitative comparison of the distribution of the stars in the MSTO
region with those in the simulations that incorporate two SSPs with an
age difference as indicated by the isochrone fitting shows for the
first time that the observed CMD of NGC\,1846 can be better explained
by a population with a distribution of ages rather than by two discrete
SSPs as suggested before. 

Finally, we compare predictions of various formation scenarios with
the observed properties of NGC~1846 described above. Scenarios where
some fraction of the stars originated from outside the cluster,
whether because NGC~1846 is the product of a binary cluster merger 
or a merger of a star cluster with a giant molecular cloud, or
because the star cluster accreted older field stars during the formation
process, do not appear to easily match all observed properties. A scenario
where a significant fraction of the stars 
form later within the cluster itself out of gas enriched by stars of
the earliest population(s) that feature slow stellar winds 
appears more consistent with the observed stellar radial density
distributions and the age range implied by the width of the MSTO
region. Viable sources of the enriched material are thought to include
fast-rotating massive 
stars and intermediate-mass AGB stars. The latter have ages similar to 
the observed range in ages across the MSTO region. Detailed abundance
ratios from high-resolution spectroscopy of individual cluster stars
should be very useful in constraining the possible source(s). 

\paragraph*{Acknowledgments.}~We are very grateful to Jay Anderson for his
support and help in using his {\it ePSF}-related programs. 
We acknowledge the helpful comments and suggestions of the anonymous
  referee. THP gratefully
acknowledges support from the National Research Council of Canada in the form
of a Plaskett Research Fellowship. Support for {\it
  HST\/} Program GO-10595  was provided by NASA through a grant from the Space
Telescope Science Institute, which is operated by the Association of
Universities for Research in Astronomy, Inc., under NASA contract
NAS5--26555. We acknowledge the use of the $R$ Language for Statistical
Computing, see http://www.R-project.org.


\begin{thebibliography}{}
\bibitem[Anderson \& King(2000)]{andkin00}
Anderson, J., \& King, I. R. 2000, \pasp,  112,  1360
\bibitem[Anderson \& King(2006)]{andkin06}
Anderson, J., \& King, I. R. 2006, ``PSFs, Photometry, and Astrometry for the
 ACS/WFC'', ACS Instrument Science Report 2006-01 (Baltimore:STScI)
\bibitem[Anderson et al.(2008a)]{ande+08a}
Anderson, J., Sarajedini, A., Bedin, L. R., King, I. R., Piotto, G., Reid, 
 I. N., Siegel, M., Majewski, S. R., et al.\ 2008, \aj, 135, 2055
\bibitem[Anderson et al.(2008b)]{ande+08b}
Anderson, J., King, I. R., Richer, H. B., Fahlman, G. G., Hansen, B. M. S.,
 Hurley, J., Kalirai, J. S., Rich, R. M., et al.\ 2008, \aj, 135, 2114
\bibitem[Bastian \& Goodman(2006)]{basgoo06}
Bastian, N., \& Goodman, S. P. 2006, \mnras, 369, L9
\bibitem[Bedin et al.(2004)]{bedi+04}
Bedin, L.~R., Piotto, G., Anderson, J., Cassisi, S., King, I.~R., Momany, Y., 
 \& Carraro, G.\ 2004, \apjl, 605, L125 
\bibitem[Bedin et al.(2005)]{bedin+05}
Bedin, L. R., Cassisi, S., Castelli, F., Piotto, G., Anderson, J., Salaris,
 M., Momany,  Y., \& Pietrinferni, A.\ 2005, \mnras, 357, 1038 
\bibitem[Bekki \& Mackey(2009)]{bekmac09}
Bekki, K., \& Mackey, A. D.\ 2009, \mnras, 
 doi:10.1111/j.1365-2966.2008.14320.x 
\bibitem[Bertelli et al.(2003)]{bert+03} 
Bertelli, G., Nasi, E., Girardi, L., Chiosi, C. Zoccali, M., \& Gallart, C.\ 
 2003, \aj, 125, 770
\bibitem[Bhatia \& Hatzidimitriou(1988)]{bhahat88}
Bhatia, R., \& Hatzidimitriou, D.\ 19
\bibitem[Briley et al.(2002)]{bril+02}
Briley, M. M., Cohen. J. G., \& Stetson, P. B.\ 2002, \apj, 579, L17
\bibitem[Briley et al.(2004)]{bril+04}
Briley, M. M., Harbeck, D., Smith, G. H., \& Grebel, E. K.\ 2004, \aj, 127,
 1588 
\bibitem[Bruzual \& Charlot(2003)]{bc03} 
Bruzual, G. A., \& Charlot, S., 2003, \mnras, 344, 1000
\bibitem[Cannon et al.(1998)]{cann+98}
Cannon, R. D., Croke, B. F. W., Bell, R. A., Hesser, J. E., \& Stathakis,
 R. A. 1998, \mnras, 298, 601
\bibitem[Cardelli et al.(1989)Cardelli, Clayton, \& Mathis]{card+89}
Cardelli, J. A., Clayton, G. C., \& Mathis, J. S. 1989, \apj, 345, 245
\bibitem[Carretta et al.(2006)]{carr+06}
Carretta, E., Bragaglia, A,. Gratton, R. G., Leone, F., Recio-Blanco, A., \&
 Lucatello, S.\ 2006, \aap, 450, 523
\bibitem[Castelli \& Kurucz(2003)]{caskur03}
Castelli, F., \& Kurucz, R. L., 2003, Modelling of Stellar
 Atmospheres, ed.\ N.\ Piskunov, W.\ W.\ Weiss, \& D.\ F.\ Gray (San
 Francisco: ASP), A20 
\bibitem[Cohen et al.(2005)]{cohe+05}
Cohen, J. G., Briley, M. M., \& Stetson, P. B.\ 2005, \aj, 130, 1177
\bibitem[Decressin et al.(2007a)]{decr+07}
Decressin, T., Meynet, G., Charbonnel, C., Prantzos, N., \& Ekstr\'om, S.\
 2007a, \aap, 464, 1029
\bibitem[Denissenkov \& VandenBerg(2003)]{denvdb03}
Denissenkov, P. A., \& Vandenberg, D. A., 1003, \apj, 593, 509
\bibitem[D'Ercole et al.(2008)]{derc+08}
D'Ercole, A., Vesperini, E., D'Antona, F., McMillan, S. L. W., \& Recchi, S.\
 2008, \mnras, doi:10.1111/j.1365-2966.2008.13915.x
\bibitem[Dolphin(2000)]{dolp00}
Dolphin, A. E. 2000, \pasp, 112, 1383
\bibitem[Dotter(2008)]{Dotter08} Dotter, A.\ 2008, \apjl, 687, 
L21
\bibitem[Dotter et al.(2008)]{dott+08}
Dotter, A., Chaboyer, B., Jevremovi\'c, D., Kostov, V., Baron, E., \&
 Ferguson, J. W.\ 2008, \apjs, 178, 89
\bibitem[Efremov \& Elmegreen(1998)]{efrelm98}
Efremov, Y. N., \& Elmegreen, B. G.\ 1998, \mnras, 299, 588
\bibitem[Elson \& Fall(1985)]{elsfal85}
Elson, R. A. W., \& Fall, S. M.\ 1985, \apj, 299, 211
\bibitem[Elson \& Fall(1988)]{elsfal88}
Elson, R. A. W., \& Fall, S. M.\ 1988, \aj, 96, 1383
\bibitem[Fellhauer et al.(2006)]{fell+06}
Fellhauer, M., Kroupa, P., \& Evans, N. W.\ 2006, \mnras, 372, 338
\bibitem[Fruchter \& Hook(2002)]{fruchook02}
Fruchter, A. S., \& Hook, R. N. 2002, \pasp, 114, 792
\bibitem[Fujimoto \& Kumai(1997)]{fujkum97}
Fujimoto, M., \& Kumai, Y.\ 1997, \aj, 113, 249
\bibitem[Fusi Pecci et al.(1990)]{fusi+90}
Fusi Pecci, F., Ferraro, F. R., Crocker, D. A., Rood, R. T., \& Buonanno, R.\ 
 1990, \aap, 238, 95
\bibitem[Goudfrooij et al.(2006)]{goud+06}
Goudfrooij, P., Bohlin, R. C., Ma\'{\i}z-Apell\'aniz, J., \& Kimble, R. A.,
 2006, \pasp, 848, 1455
\bibitem[Goudfrooij et al.(2007)]{goud+07}
Goudfrooij, P., Schweizer, F., Gilmore, D., \& Whitmore, B. C. 2007,
 \aj, 133, 2737
\bibitem[Girardi et al.(1995)]{gira+95}
Girardi, L., Chiosi, C., Bertelli, G., \& Bressan, A. 1995, \aap, 298, 87
\bibitem[Girardi et al.(2000)]{gira+00}
Girardi, L., Bressan, A., Bertelli, G., \& Chiosi, C. 2000, A\&AS, 141, 371 
\bibitem[Girardi et al.(2008)]{gira+08}
Girardi, L., Dalcanton, J., Williams, B., de Jong, R. S., Gallart, C., et al.\
 2008, \pasp, 120, 583
\bibitem[Gratton et al.(2001)]{grat+01}
Gratton, R. G., et al.\ 2001, \aap, 369, 87
\bibitem[Hauschildt et al.(1999)]{haus+99}
Hauschildt, P. H., Allard, F., Ferguson, J., Baron, E., \& Alexander,
 D.\ 1999b, \apj, 525, 871 
\bibitem[Hilker \& Richtler(2000)]{hilric00}
Hilker, M., \& Richtler, T.\ 2000, \aap, 362, 895
\bibitem[Iben(1968)]{iben68}
Iben, I., Jr.\ 1968, \nat, 220, 143
\bibitem[Kerber et al.(2007)Kerber, Santiago, \& Brocato]{kerb+07}
Kerber, L. O., Santiago, B. X., \& Brocato, E., 2007, \aap, 462, 139
\bibitem[King(1962)]{king62}
King, I. 1962, \aj, 67, 471 
\bibitem[Kozhurina-Platais et al.(2007)Kozhurina-Platais, Goudfrooij, \& Puzia]{kozh+07}
Kozhurina-Platais, V., Goudfrooij, P., \& Puzia, T. H. 2007, ACS Instrument
 Science Report 2007-04 (Baltimore: STScI)
\bibitem[Koekemoer et al.(2003)]{koek+03} Koekemoer, A. M., Fruchter,
  A. S., Hook, R. N., \& Hack, W.\ 2003, in ``2002 HST
  Calibration Workshop'', eds.\ S. Arribas, A. Koekemoer, \& B. Whitmore
  (Baltimore: STScI), 337
\bibitem[Lee(1991)]{Lee91} Lee, Y.-W.\ 1991, \apjl, 373, L43
\bibitem[Lebzelter \& Wood(2007)]{lebwoo07} 
Lebzelter, T., \& Wood, P. R.\ 20076, \aap, 475, 643
\bibitem[Leon et al.(1999)]{leon+99}
Leon, S., Bergond, G., \& Vallenari, A., 1999, \aap, 344, 450
\bibitem[Mackey \& Broby Nielsen(2007)]{macbro07}
Mackey, A. D., \& Broby Nielsen, P. 2007, \mnras, 379, 151
\bibitem[Mackey et al.(2008)]{mack+08}
Mackey, A. D., Broby Nielsen. P., Ferguson, A. M. N., \& Richardson,
 J. C. 2008, \apj, 681, L17
\bibitem[Maraston(2005)]{mara05}
Maraston, C., 2005, \mnras, 362, 799
\bibitem[Maraston et al.(2008)]{mara+08}
Maraston, C., Str\"omb\"ack, G., Thomas, D., Wake, D. A., \& Nichol, D. 2008,
 \mnras, 394, L107
\bibitem[Marigo et al.(2008)]{mari+08}
Marigo, P., Girardi, L., Bressan, A., Groenewegen, M. A. T., Silva, L., \&
 Granato, G. L. 2008, \aap, 482. 883
\bibitem[Meylan \& Heggie(1997)]{mayheg97}
Meylan, G., \& Heggie, D. C. 1997, \aapr, 8, 1
\bibitem[Mighell et al.(1998)Mighell, Sarajedini, \& French]{migh+98}
Mighell, K. J., Sarajedini, A., \& French, R. S. 1998, \aj, 116, 2395
\bibitem[Mucciarelli et al.(2007)]{mucc+07}
Mucciarelli, A., Ferraro, F. R., Origlia, L., \& Fusi Pecci, F. 2007, \aj, 133, 
	2053
\bibitem[Milone et al.(2008a)]{milo+08a}
Milone, A. P., et al.\ 2008a, \apj, 673, 241
\bibitem[Milone et al.(2008b)]{milo+08b}
Milone, A. P., Bedin, L. R., Piotto, G., \& Anderson, J.\ 2008b, \aap, in
 press (arXiv:0810.2558v1) 
\bibitem[Norris et al.(1996)Norris, Freeman, \& Mighell]{norr+96}
Norris, J. E., Freeman, K. C., \& Mighell, K. J.\ 1996, \apj, 462, 241
\bibitem[Pessev et al.(2006)]{pess+06}  
Pessev, P. M., Goudfrooij, P., Puzia, T. H., \& Chandar, R.\ 2006, \aj, 132, 781
\bibitem[Pessev et al.(2008)]{pess+08} 
Pessev, P. M., Goudfrooij, P., Puzia, T. H., \& Chandar, R.\ 2008, \mnras, 385, 1535
\bibitem[Pflamm-Altenburg \& Kroupa(2007)]{pflkro07}
Pflamm-Altenburg, J., \& Kroupa, P.\ 2007, \mnras, 375, 855
\bibitem[Pickles(1998)]{pick98}
Pickles, A. J. 1998, \pasp, 110, 863
\bibitem[Pietrinferni et al.(2004)]{piet+04}
Pietrinferni, A., Cassisi, S., Salaris, M., \& Castelli, F. 2004, \apj, 612,
 168
\bibitem[Pietrinferni et al.(2006)]{piet+06}
Pietrinferni, A., Cassisi, S., Salaris, M., \& Castelli, F. 2006, \apj, 642, 
 797
\bibitem[Piotto(2008)]{piot08}
Piotto, G., 2008, Mem. S. A. It., 79, 3
\bibitem[Piotto et al.(2007)]{piot+07}
Piotto, G., et al.\ 2007, \apjl, 661, L53
\bibitem[Recchi \& Danziger(1995)]{recdan05}
Recchi, S., \& Danziger, I. J. 2005, \aap, 436, 145
\bibitem[Riess \& Mack(2004)]{riesmack04}
Riess, A., \& Mack, J. 2004, ACS Instrument Science Report 2004-06
 (Baltimore: STScI)
\bibitem[Saslaw(1985)]{sasl85}
Saslaw, W. C. 1985, Gravitational Physics of Stellar and Galactic Systems 
 (Cambridge: Cambridge University Press)
\bibitem[Searle et al.(1980)Searle, Wilkinson, \& Bagnuolo]{swb80}
Searle, L., Wilkinson, A., \& Bagnuolo, W. G.\ 1980, \apj, 239, 803
\bibitem[Silverman(1986)]{silv86} 
Silverman, B. W. 1986, in {\it Density Estimation for Statistics and
  Data Analysis}, Chap and Hall/CRC Press, Inc. 
\bibitem[Sirianni et al.(2005)]{siri+05} Sirianni, M., Jee, M. J.,
  Ben\'{\i}tez, N., Blakeslee, J. P., Martel, A. R., Meurer, G., Clampin,
  M., De Marchi, G., et al.\ 2005, \pasp, 117, 1049. 
\bibitem[Spitzer(1987)]{spit87}
Spitzer, L. Jr. 1987, Dynamical Evolution of Globular Clusters (Princeton:
 Princeton University Press)\
\bibitem[Suntzeff \& Smith(1991)]{sunsmi91}
Suntzeff, N. B., \& Smith, V. 1991, \apj, 381, 160
\bibitem[Theis(2002)]{thei02}
Theis, Ch.\ 2002, \apss, 281, 97
\bibitem[Thomas(1967)]{thom67}
Thomas, H.-C. 1967, ZAp, 67, 420
\bibitem[van den Bergh(2004)]{vdb04}
van den Bergh, S.\ 2004, \aj, 127, 897
\bibitem[van der Wel et al.(2006)]{vdwel+06}
van der Wel, A., Franx, M., van Dokkum, P. G., Huang, J., Rix, H.-W., \&
 Illingworth, G. D. 2006, \apj, 2006, L21
\bibitem[Ventura et al.(2001)]{vent+01}
Ventura, P., D'Antona, F., Mazzitelli, I., \& Gratton, R. G.\ 2001, \apjl,
 550, L65
\bibitem[Ventura et al.(2002)]{vent+02}
Ventura, P., D'Antona, F., Mazzitelli, I.\ 2002, \aap, 393, 215
\bibitem[Ventura \& D'Antona(2008)]{vendan08}
Ventura, P., \& D'Antona, F.\ 2008, \mnras, 385, 2034
\bibitem[Villanova et al.(2007)]{vill+07}
Villanova, S., et al.\ 2008, \apj, 663, 296
\end{thebibliography}
\end{document}